\documentclass[11pt,aps,pra,superscriptaddress,showpacs,floatfix]{revtex4}
\usepackage{graphicx,psfrag}
\usepackage{subfigure}
\usepackage{amsmath,amssymb}
\usepackage{type1cm}
\usepackage{bm}
\usepackage{comment}
\usepackage{mathptmx}
\usepackage{wrapfig}

\newcommand{\Nt}{\mbox{${\tilde N}_T$}}

\newcommand{\Et}{\mbox{${\tilde E}_T$}}

\newcommand{\tlX}{\mbox{${\tilde X}$}}

\newcommand{\tlT}{\mbox{${\tilde T}$}}

\newcommand{\vx}{\bm{x}}

\newcommand{\vp}{\mbox{$\bm{p}$}}

\newcommand{\vbr}{\mbox{$\bm{r}$}}

\newcommand{\vlambda}{\mbox{$\bm{\lambda}$}}
\newcommand{\tvlamb}{\mbox{$^t{\bm{\lambda}}$}}
\newcommand{\dvlamb}{\mbox{$\bm{\dot \lambda}$}}
\newcommand{\tdvlamb}{\mbox{$^t{\bm{\dot \lambda}}$}}

\newcommand{\Ybfst}{$^{170}$Yb-$^{171}$Yb\,}
\newcommand{\Ybsnd}{$^{170}$Yb-$^{173}$Yb\,}
\newcommand{\Ybthd}{$^{174}$Yb-$^{173}$Yb\,}

\begin{document}

\title{Longitudinal and Transverse Breathing Oscillations 
\\in Bose - Fermi Mixtures of Yb atoms at Zero Temperature 
\\in the Largely Prolate Deformed Traps}

\author{Tomoyuki~Maruyama}
\affiliation{College of Bioresource Sciences,
Nihon University,
Fujisawa 252-0880, Japan}
\affiliation{Advanced Science Research Center,
Japan Atomic Energy Agency, Tokai 319-1195, Japan}
\affiliation{National Astronomical Observatory of Japan, 
2-21-1 Osawa, Mitaka, Tokyo 181-8588, Japan}

\author{Hiroyuki Yabu}
\affiliation{
Department of Physics,
Ritsumeikan University, Kusatsu 525-8577, Japan}

\date{\today}

\begin{abstract}
We study the breathing oscillations in bose-fermi mixtures
of Yb isotopes in the largely prolate deformed trap, 
which are realized by Kyoto group.
We take three combinations of the Yb isotopes,
\Ybfst, \Ybsnd and \Ybthd, whose boson-fermion interactions are weakly
repulsive, strongly attractive and strongly repulsive.
The collective oscillations in the deformed trap 
are calculated in the dynamical time-development approach, 
which is formulated with the time-dependent Gross-Pitaevskii 
and the Vlasov equations. 
We analyze the results 
with the intrinsic oscillation modes of the deformed system, 
obtained in the scaling method, 
and show that the damping and forced-oscillation effects 
of the intrinsic modes explain time-variation behaviors of oscillations, 
especially, in the fermion transverse mode.
\end{abstract}


\maketitle

\section{Introduction}
\label{intr}

In this decade,
there have been significant progresses 
in the production of ultracold gases, 
which realize the Bose-Einstein condensates (BEC) \cite{nobel,Dalfovo,becth,Andersen}, 
two boson mixtures  \cite{Truscott,B-B}, 
degenerate atomic Fermi gases \cite{ferG},
and Bose-Fermi (BF) mixing gases \cite{Truscott,Schreck,BferM,Modugno}.   
In particular the BF mixtures attract physical interest 
as a typical example in which particles 
obeying different quantum statistics are intermingled.
Using the system we have a very big opportunity to obtain
various new knowledge of many-body quantum systems 
because we can make a large variety of combinations of 
atomic species and control the atomic interactions 
using the Feshbach resonance \cite{Fesh}.

Theoretical studies of the BF mixtures have been done on static 
properties \cite{Molmer,Amoruso,MOSY,Bijlsma,Vichi,Vichi1}, 
the phase diagram and phase
separation \cite{Nygaard,Yi,Viverit,Capuzzi02}, 
stability  \cite{MSY1,Roth,Capuzzi03} and
collective excitations
\cite{MSY,Minguzzi,MOSY,zeros,Yip,sogo,tomoBF,monoEX,tbDPL,QPBF}.

Recently Kyoto university group has performed researches on the trapped atomic
gases of the Yb isotopes:
the BEC \cite{Takasu} and the Fermi-degeneracy \cite{Fukuhara1}.  
They also succeeded in realizing
the BF mixtures of the isotopes. 
The Yb consists of many kinds of isotopes--- 
five bosons (${}^{168,170,172,174,176}$Yb) 
and two fermions (${}^{171,173}$Yb), 
which give a variety of combinations in the BF mixtures.
The scattering lengths of the boson-fermion interactions have been determined
experimentally by the group \cite{Fukuhara1,Fukuhara2,Kitagawa,Fukuhara3,Fukuhara-M}, 
and the experimental studies of the ground state properties and the collective oscillations 
of the BF mixtures is now under progressing.

In many characteristic properties, 
the spectrum of the collective excitations is
an important diagnostic signal of these systems;
they commonly appear in many-particle systems
and are often sensitive to the inter-particle interaction and the structure of the
ground and excited states.  

We have constructed a dynamical
approach to solve time-developments of the oscillation states 
with the time-dependent Gross-Pitaevskii (TDGP) and Vlasov equations, 
and have studied the monopole  \cite{tomoBF,monoEX} 
and dipole oscillations  \cite{tbDPL} of the BF mixtures.
In these works, 
we have found that 
the oscillational motions of the BF mixtures include various modes 
such as the boson and fermion intrinsic and 
the forced oscillation modes, 
and that the intrinsic frequencies of these modes obtained in the dynamical calculation are different 
from those in the sum-rule \cite{MOSY,MSY}
and the scaling \cite{scal,scalDP,scalTF,scalTF2} approaches, 
which are approximations of the random phase approximation (RPA) \cite{zeros,sogo}.
Furthermore, the dynamical calculations are consistent with the results in RPA only in early stage of time, 
but show discrepancies in later stage of time, 
and, as the boson-fermion interaction becomes stronger, 
the discrepancies come to appear earlier in oscillation.
Thus the BF mixtures show new dynamical properties different from the other finite many-body systems 
such as atomic nuclei.

In this paper, 
we take three combinations of the Yb isotopes,
\Ybfst, \Ybsnd and \Ybthd, 
where the boson-fermion interactions are weakly repulsive, strongly attractive and strongly repulsive.
In the previous paper \cite{QPBF}, 
we have calculated the quadrupole oscillation of these mixtures
in the spherical trap, 
and compared the results with those obtained in RPA.
The RPA gives small-amplitude intrinsic modes of the mixtures
and describes the whole oscillations
as linear combinations of these intrinsic modes.
When the boson-fermion interaction is weak and/or the amplitude is small,
the linearly-combined oscillations can be good approximation,
and the RPA gives the results consistent 
with those in the dynamical approach. 
However, when the boson-fermion interaction is strong or the amplitude is large,  
the oscillations of the BF mixtures obtained in the dynamical approach 
show very different behaviors from those obtained in the RPA.
For example, in the strongly-attractive boson-fermion interaction, 
the fermi gas overflows from the boson occupation region and makes large expansion, 
which leads to a monopole oscillation mode,   
and, in the case of the strongly-repulsive interaction, 
the fermion intrinsic-mode shows rapid damping 
so that the fermi gas comes to co-oscillate with the bose gas \cite{QPBF}.

In actual experiments,
the largely-deformed trap of axial-symmetry is used, 
and the oscillation behaviors should be different from those in the spherical trap.
The monopole and quadrupole oscillations, 
which are coupled in axial-symmetric traps, 
are recombined into two modes oscillating along the symmetric axial and the
transverse directions.
Such behaviors are observed in the two-component fermi-gas oscillations
\cite{scalTF2}, but no studies have been done for the BF mixture.

In this paper, 
we investigate the breathing oscillations of the BF mixtures in
the largely deformed system of the prolate shape.
Then we examine oscillation behaviors along the longitudinal and
transverse directions separately, 
which are  the symmetry axis and 
the directions perpendicular to it, respectively.     

In the next section, 
we explain a microscopic model of the BF mixtures,
and examine the intrinsic collective modes and their coupling
behaviors in the scaling method.
In Sec. III, 
we show the time-development calculations 
in the TDHF+Vlasov approach
for the breathing oscillations in the BF mixture, 
and discuss their properties
on the basis of the collective modes obtained in the scaling method.
The summary of the present work is given in Sec. IV.

\newpage

\section{Collective Oscillations in  Scaling Method}
\label{Te}

Before presenting actual time-development calculations 
we consider the oscillation behaviors in the deformed system, 
and take up the collective modes  
from the macroscopic point of view.

The deformed BF mixtures have the breathing oscillations, 
which are generally coupled modes 
of the monopole and quadrupole oscillations.
The mode-structure of breathing oscillations
in the deformed system has been studied, 
for example, in two-component fermi gas \cite{scalTF2}, 
but it is not clear for the deformed BF mixtures.
Thus, in the present section, 
we should clarify the mode-structure of the breathing oscillations 
in the deformed BF mixture
using the scaling method as in Ref.~\cite{sumF};
it is a macroscopic approximation
consistent with the energy-weighted sum rules \cite{bohi}.

The scaling method describe 
the collective oscillations within the linear response regime.
However, the method has been extended in the nonlinear case 
with contributions of the nonlinear terms.
The nonlinear calculations was applied 
to the oscillations of BEC \cite{CD96}, 
which reproduced the experimental results by Ref.~\cite{Mewes96} very well. 
Also, the method was used in the study of the BEC 
with non-condensed bosons \cite{KSS96}.

\subsection{Total Energy of BF mixtures and Scaled Variables}

In this work we consider the mixture of dilute boson and 
one-component-fermion gases at zero temperature 
in an axially symmetric trapping potential,
the symmetry axis of which is chosen to be the $z$-axis. 
We assume zero-range interactions between atoms, 
and no fermion-fermion interactions. 
Using the Hartree-Fock approximation, 
the total energy is given by a functional
of the condensed boson wave function $\phi_c$, 
and the $n$-th fermion single-particle wave functions $\psi_n$:
\begin{eqnarray}
     E_T &=&\int d^3{r} \Bigg[
            -\frac{1}{2} \nabla_r \phi_c^\dagger(\vbr) 
                         \nabla_r \phi_c(\vbr)
            +\frac{1}{2} (\vbr_T^2 +\kappa_L^2 z^2) 
                          \phi_c^\dagger(\vbr) \phi_c(\vbr) 
            +\frac{g_{BB}}{2} \{ \phi_c^\dagger(\vbr) 
                                 \phi_c (\vbr)      \}^2  
\nonumber \\
         &&\qquad\quad
         +\frac{1}{2m_f} \sum_n \nabla_r \psi_n^\dagger(\vbr) 
                                 \nabla_r \psi_n (\vbr)
         +\frac{1}{2} m_f \omega_f^2 
          (\vbr_T^2 + \kappa_L^2 z^2) 
          \sum_n \psi_n^\dagger(\vbr) \psi_n (\vbr) 
\nonumber \\
         &&\qquad\quad
         +h_{BF} \phi_c^\dagger (\vbr) \phi_c (\vbr) 
          \sum_n \psi_n^\dagger (\vbr) \psi_n (\vbr) \Bigg] .
\label{etot}
\end{eqnarray}
In the above equation all the variables are dimensionless.
The spatial coordinate $\vbr$ is scaled with $(\hbar / M_B \Omega_B)^{1/2}$, 
where $M_B$ and $\Omega_B$ are the boson mass and
boson transverse trapped frequency,
$m_f$ and $\omega_f$ are the fermion-to-boson ratios of the mass 
and the trapping-potential frequency,
and $g_{BB}$ and $h_{BF}$ are 
the dimensionless coupling constants of the boson-boson and
boson-fermion interactions; the detailed definitions are given 
in Ref.~ \cite{tomoBF,tbDPL,QPBF}.

Here the scaled wave functions are introduced 
from the ground-state one-body wave functions, 
$\phi_c^{(g)}$ and $\psi^{(g)}_n$:
\begin{eqnarray}
     \phi_{\lambda}(\vbr,\tau) &=& 
          e^{i \xi_B (\vbr,\tau)} 
          e^{\lambda_{BT}(\tau) +\frac{1}{2}\lambda_{BL}(\tau)} 
          \phi_c^{(g)}(e^{\lambda_{BT}(\tau)} \vbr_T; e^{\lambda_{BL}(\tau)} z),
\label{scwfB}\\
     \psi_{\lambda,n}(\vbr,\tau) &=& 
          e^{i m_f \xi_F (\vbr,\tau)} 
          e^{\lambda_{FT}(\tau) + \frac{1}{2}\lambda_{FL}(\tau)} 
          \psi_n^{(g)}(e^{\lambda_{FT}(\tau)} \vbr_T; e^{\lambda_{FL}(\tau)} z)
\label{scwfF}
\end{eqnarray}
where $\tau$ is the dimensionless time coordinate scaled with
 $\Omega_B$, and $B$ and $F$ denote the boson and fermion, respectively.
The factor $\exp(i\xi_a)$ is the Gallilei-transformation factor,
which is necessary for the scaled wave functions to satisfy the continuum equation, 
and the phase parameters $\xi_a(\vbr,\tau)$ are given by
\begin{equation}
     \xi_a(\vbr,\tau) = \frac{1}{2} 
          \left[ {\dot \lambda}_{Ta}(\tau) \vbr_T^2  
                +{\dot \lambda}_{La}(\tau) z^2 \right] , \qquad
     (a=B,F)
\end{equation}

The variables, 
$\lambda_{BT}$, $\lambda_{BL}$, $\lambda_{FT}$, $\lambda_{FL}$, 
are the collective coordinates 
describing the boson longitudinal breathing (BLB), 
boson transverse breathing (BTB), fermion longitudinal breathing (FLB) 
and fermion transverse breathing (FTB) oscillation modes,
and ${\dot \lambda}$'s are the time-derivatives of them.
As it turns out, 
the four breathing modes are completely decoupled
in the BF mixtures of no interactions: $g_{BB} = h_{BF} =0$.

Here we give a comment on the method of introducing the collective coordinates.
In Ref.~ \cite{CD96}, they introduced the scaling parameters 
as $r_i \rightarrow \lambda^{\prime}_i r_i$.
For the ground state, $\lambda_i = 0$ in the present paper 
corresponds to $\lambda^\prime_i=1$,
and, in addition, $\lambda_i = - \infty$ corresponds to $\lambda^\prime_i=0$.  
For the small amplitude oscillations ($|\lambda_i| \ll 1$), 
the power expansion of the energy of he system with $\lambda_i$ should
make clear the formal differences of both methods in appearance,
while the calculated collective frequencies does not depend on the choice of the method in principle.

Furthermore, 
one can make a scaled wave function exact 
by introducing a time-dependent trap frequency and couplings with suitable
parameters \cite{GBD10}.
This method has been used for the analysis of 
the adiabatic expansions of BEC \cite{Campo11}.

Substituting the wave-functions (\ref{scwfB},\ref{scwfF})
into the total energy functional (\ref{etot}),
we obtain the total energy:
\begin{eqnarray}
     E_T &=& \frac{1}{2}  \int d^3 r 
             \left\{ (\nabla_r \xi_B)^2 \rho_B 
                    +e^{2 \lambda_{BT}} ( T_{B,1} + T_{B,2} )
                    +e^{2 \lambda_{BL}}  T_{B,3} \right\}
\nonumber \\
            && +\frac{1}{2} \int d^3 r 
                \left\{ e^{- 2 \lambda_{BT} } (r_1^2 + r_2^2)
                       +e^{-2 \lambda_{BL}} \kappa_L^2 z^2 )\right\} 
                     \rho_B (\vbr) 
\nonumber \\
           && +\frac{g_{BB}}{2} e^{2 \lambda_{BT} + \lambda_{BL} }
               \int d^3 r  \rho_B^2 (\vbr) , 
\nonumber \\
           && +\frac{1}{2 m_f} \int d^3 r 
               \left\{ m_f^2 (\nabla_r \xi_F)^2 \rho_F 
                      +e^{2 \lambda_{FT}} ( T_{F,1} + T_{F,2} )
                      +e^{2 \lambda_{FL}}  T_{F,3} \right\}
\nonumber \\
           && +\frac{1}{2} m_f \omega_f^2 \int d^3 r 
               \left\{ e^{- 2 \lambda_{FT} } (r_1^2 + r_2^2)
                      +e^{-2 \lambda_{FT}} \kappa_L^2 z^2 ) \right\} 
               \rho_F (\vbr) , 
\nonumber \\
          && +h_{BF} e^{2 \lambda_{BT} 
             +\lambda_{BL} 
             +2 \lambda_{FT}
             +\lambda_{FL}} 
              \int d^3 r\, 
              \rho_B(e^{\lambda_{BT}} \vbr_T; e^{\lambda_{BL}} z)
              \rho_F(e^{\lambda_{FT}} \vbr_T; e^{\lambda_{FL}} z),
\label{etotc}
\end{eqnarray}
where
\begin{eqnarray}
     T_{B,i} &=& \int d^3 r  \left| 
\frac{\partial}{\partial r_i} \phi (\vbr) \right|^2,
     \qquad (i=1,2,3)
\\
     T_{F,i} &=& \sum_n \int d^3{r} 
       \left| \frac{\partial}{\partial r_i} \psi_n (\vbr) \right|^2, 
     \qquad (i=1,2,3)
\\
     \rho_B (\vbr) &=& N_b |\phi_c (\vbr)|^2, \quad
     \rho_F (\vbr) ~=~ \sum^{occ}_{n} |\psi_n (\vbr)|^2.
\end{eqnarray} 

In order to obtain the ground state of the system 
we use the Thomas-Fermi (TF) approximation:
\begin{equation}
     T_B (\vbr) =0,  \qquad
     T_{F,1} (\vbr) =T_{F,2} (\vbr)  
                      =T_{F,3} (\vbr) 
                      =\frac{1}{5} (6 \pi^2)^{\frac{2}{3}} 
                       \left[\rho_F (\vbr)\right]^{\frac{5}{3}} .
\end{equation}
It should be noted that the momentum distribution is spherically-symmetric.
In order to simplify the analytic calculation, 
we use the dimensionless constants and variables:
\begin{eqnarray}
     h &= \frac{1}{m_f \omega_f^2} \frac{h_{BF}}{g_{BB}}, \quad
     & \vx = \frac{m_f^4 \omega_f^5 g_{BB}}{3 \pi^2} ( r_1, r_2, \kappa_L z ), 
\\
     n_B &= \frac{2 m_f^8 \omega_f^{10} g_{BB}^3}{9 \pi^4} \rho_B, \quad
     & n_F = \frac{2 m_f^9 \omega_f^{12} g_{BB}^3}{9 \pi^4} \rho_F,
\end{eqnarray}
with which the scaled dimensionless total energy is given by:
\begin{eqnarray}
     \Et^{(0)} &=& 
          \frac{4 \kappa_L m_f^{28} \omega_f^{35} g_{BB}^8 }{ 
                3^7 \pi^{14}} 
          E_T(\vlambda=0,{\dot \vlambda}=0)
\nonumber \\
          &=& \int d^3{x} 
              \Bigg\{ x^2 n_B  
                     +\frac{1}{2} n_B^2 
                     +\frac{3}{5} n_F^{\frac{5}{3}} 
                     +x^2 n_F 
                     +h n_B n_F \Bigg\} ,
\label{etotTFsg}
\end{eqnarray}
where $x^2 = |\vx|^2$. 
After the scaling transformation, the ground state depends
only on the three variables, $e_B$, $e_F$ and $h$, and 
the dependence of the deformation parameter $\kappa_L$, 
the boson-boson coupling $g_{BB}$, 
the fermion mass $m_f$ are eliminated.

From the variations of the total energy with respect to the densities, 
$\delta {\tilde E}^{(0)}/\delta n_B = 0$ and 
$\delta {\tilde E}^{(0)}/\delta n_F = 0$ ,
we obtain the TF equations:
\begin{equation}
     n_B + h n_F = e_B - x^2, \quad
     n_F^{\frac{2}{3}} + h n_B = e_F - x^2.
\label{TFeq}
\end{equation}
where $e_B$ and $e_F$ are the scaled chemical potentials for boson and fermion,
respectively, 
which are written with the chemical potentials for boson $\mu_B$
and fermion $\mu_F$ as 
\begin{eqnarray}
 e_B &=& \frac{2 m_f^8 \omega_f^{10} g_{BB}^2}{9 \pi^4} \mu_B, \\
 e_F &=& \frac{2 m_f^7 \omega_f^{8} g_{BB}^2}{9 \pi^4} \mu_F .
\end{eqnarray}

The stability condition of the TF solution is obtained 
from the second-order variation of the total energy:
\begin{equation}
     \frac{ \delta^2 {\tilde E}^{(0)} }{ \delta n_B^2 }
     \frac{ \delta^2 {\tilde E}^{(0)} }{ \delta n_F^2 }
     -\left( \frac{ \delta^2 {\tilde E}^{(0)} }{ 
                    \delta n_B \delta n_F} \right)^2 
     =\frac{2}{3} n_F^{-\frac{1}{3}} - h^2 > 0,
\end{equation}
Using (\ref{etotTFsg}), 
it gives the upper limit of the fermion density:
\begin{equation}
n_F  < \frac{8}{27 h^6} = \left(\frac{2}{3h^2} \right)^3.  
\label{StTF}
\end{equation}

Furthermore, from Eqs.~(\ref{TFeq}), 
the spatial derivative of $n_F$ is related with $n_F$ itself 
in the TF approximation:
\begin{equation}
\frac{\partial n_F}{\partial x^2}
= \frac{h - 1}{ \frac{2}{3} n_F^{-\frac{1}{3}} - h^2 }.
\label{dFdx}
\end{equation}
Thus it proves that $\partial n_F/\partial x^2 < 0$ when $h < 1$,
and $\partial n_F/\partial x^2 > 0$ when $h > 1$.

\subsection{Collective Oscillations}

Now we study the collective oscillation in the scaling method.
With the scaling variables, 
the total energy for collective oscillations (\ref{etotc}) 
is written as
\begin{eqnarray}
      {\tilde E}_T &=&  
          \frac{1}{2} {\tilde X}_B \left\{ 
               \frac{2 e^{- 2 \lambda_{BT}}}{3}{\dot \lambda}^2_{BT}
             +\frac{e^{-2 \lambda_{BL}}}{3 \kappa_L^2}{\dot \lambda}^2_{BL}
          \right\}
\nonumber \\
     && +\frac{2 e^{-2 \lambda_{BT}} 
           +e^{-2\lambda_{BL}} }{3} {\tilde X}_B
	    + e^{2 \lambda_{BT} + \lambda_{BL}} V_{bb}
\nonumber \\
     && +\frac{1}{2} {\tilde X}_F \omega_f^2 \left\{ 
             \frac{2 e^{- 2 \lambda_{FT}}}{3}{\dot \lambda}^2_{FT}
           +\frac{e^{-2 \lambda_{FL}}}{3 \kappa_L^2}{\dot \lambda}^2_{FL}
          \right\}
\nonumber \\
     && +\frac{2 e^{2 \lambda_{FT}} 
           +e^{2 \lambda_{FL}}}{3} {\tlT}_F
           +\frac{2 e^{-2 \lambda_{FT}} 
           +e^{-2\lambda_{FL}}}{3} {\tilde X}_F
	    +V_{bf}
\label{etotTFs}
\end{eqnarray}
where
\begin{eqnarray}
     \tlT_{F}   &=& \frac{3}{5} \int d^3{x} \,  
                          n_{F}^{\frac{5}{3}}(\vx) \\
    X_{B,F} &=& \int d^3{x}\, x^2 n_{B,F}(\vx) \\
    V_{bb}  &=& \int d^3{x}\, n_{B}^2(\vx) \\
    V_{bf} &=& e^{ 2\lambda_{BT} +\lambda_{BL} 
                             +2\lambda_{FT} +\lambda_{FL}} 
\nonumber \\
          && \times h \int d^3{r}\, 
                    n_B(e^{\lambda_{BT}}\vx_T,e^{\lambda_{BL}} x_3 )
                    n_F(e^{\lambda_{FT}}\vx_T,e^{\lambda_{FL}} x_3 )
\end{eqnarray}

Using the vector notation for the collective variables,
$\tvlamb=(\lambda_{BT}, \lambda_{BL}, \lambda_{FT}, \lambda_{FL})$, 
and expanding the total energy 
to the order of $O(\lambda^2$), 
we obtain the oscillation energy of the system:
\begin{equation}
     \Delta {\tilde E}_T \equiv {\tilde E}_T - {\tilde E}^{(0)}_T 
                \approx \frac{1}{2} \tdvlamb B \dvlamb
                            +\frac{1}{2} \tvlamb C \vlambda.
\label{OsE}
\end{equation}
The matrices $B$ and $C$ are defined by
\begin{equation}
 B = \left( \begin{array}{cccc} 
                 \frac{4 X_B}{3} & 0 & 0 & 0 \\
                0 & \frac{2}{3 \kappa_L^2} X_B &  0 & 0 \\
                0 & 0 &  \frac{4}{3 \omega_f^2}  X_F  &  0 \\
                0 & 0 &  0 & \frac{2}{3 \kappa_L^2 \omega_f^2}  X_F  \\
                    \end{array} \right) ,
\label{MTmx}
\end{equation}
\begin{eqnarray}
    C &=& \left( \begin{array}{cccc} 
                  \frac{8}{3}X_B + 4 V_{bb}  & 2 V_{bb} & 0 & 0 \\
                  2 V_{bb} &  \frac{4}{3}X_B +  V_{bb}  & 0 & 0 \\
                  0 & 0 & \frac{8}{3}(\tlT_F + X_F) & 0 \\
                  0 & 0 & 0 & \frac{4}{3}(\tlT_F + X_F) \\
                   \end{array} \right) 
\nonumber \\
&&  + \left( \begin{array}{cccc} 
     ~- \frac{4}{3}V_1 - \frac{8}{15}V_3 ~& 
     ~ - \frac{2}{3}V_1 - \frac{2}{15}V_3~ &
     ~ \frac{8}{15} V_3 &  \frac{2}{15} V_3 ~ \\
     ~ - \frac{2}{3} V_1 - \frac{2}{15} V_3 ~ & 
     ~ - \frac{1}{3} V_1 - \frac{1}{5} V_3 ~ & 
     ~ \frac{2}{15} V_3 ~ & \frac{1}{5} V_3 ~ \\
     ~ \frac{8}{15} V_3 ~ & ~\frac{2}{15} V_3 ~ & 
     ~ -\frac{4}{3} V_2 - \frac{8}{15} V_3 ~ &  
     ~ -\frac{2}{3} V_2 - \frac{2}{15} V_3 ~\\
     ~\frac{2}{15} V_3 & \frac{1}{5} V_3 ~ & ~ 
     -\frac{2}{3} V_2 - \frac{2}{15} V_3 ~
     &  - \frac{1}{3} V_2 - \frac{1}{5} V_3 \\
 \end{array} \right)  , 
\label{RFmx}
\end{eqnarray}
where
\begin{eqnarray}
     V_{1} &=& h \int d^3{x}\,
               x\, n_B \frac{\partial n_F}{\partial x} ,
\\
     V_{2} &=& h \int d^3{x}\, 
               x\, n_F \frac{\partial n_B}{\partial x}, 
\\
     V_{3} &=& h \int d^3{x}\,
                x^2 \frac{\partial n_B}{\partial x} 
                    \frac{\partial n_F}{\partial x}.
\label{V3}
\end{eqnarray}
The stability conditions of the breathing-oscillation states 
are obtained from
$\partial {\tilde E}_T / \lambda_\alpha = 0$:
\begin{eqnarray}
     -2 X_B + 3 V_{bb} -  V_{1} &=& 0
\label{Bstb}\\
     2 \tlT_F - 2 X_F - V_{2} &=& 0.
\label{Fstb}
\end{eqnarray}

\begin{wrapfigure}{r}{8.cm}
\vspace*{-0.5cm}
\begin{center}
{\includegraphics[scale=0.55]{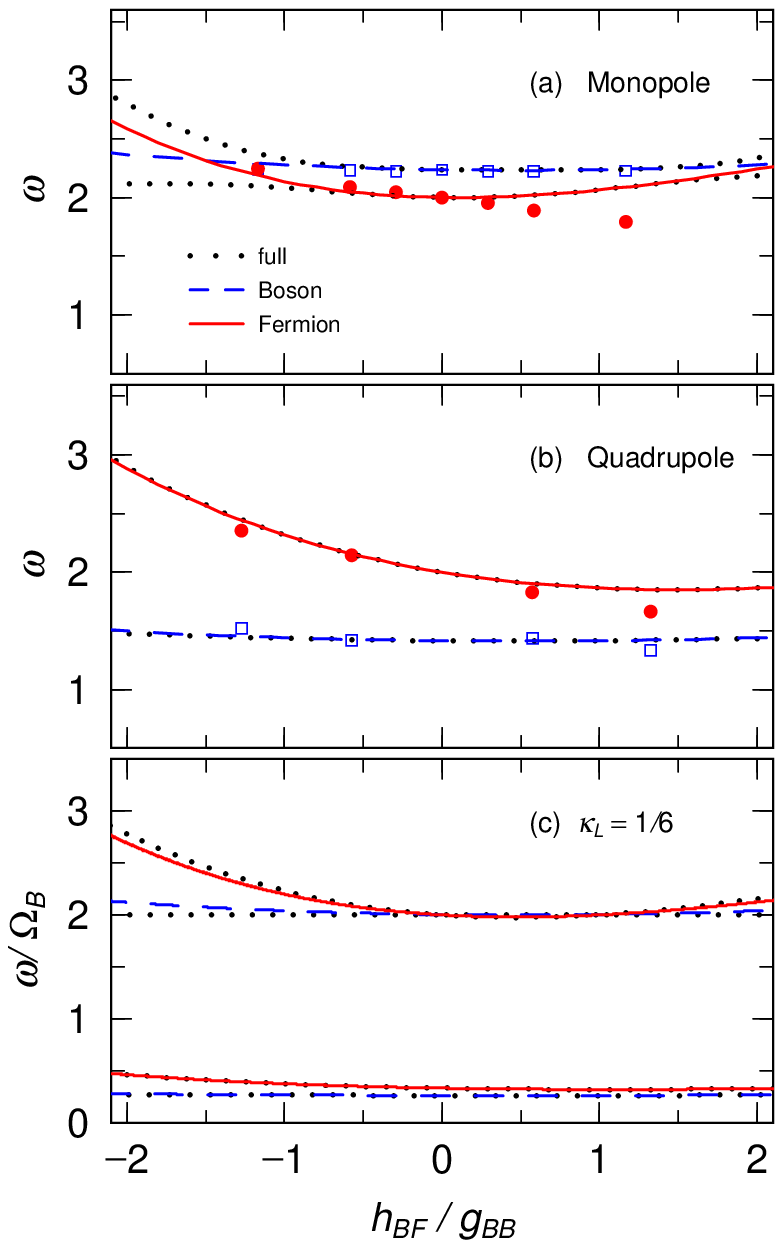}}
\caption{\small (Color online)
The frequencies of the monopole (a) and the quadrupole oscillations (b) 
in the spherically-symmetric system ($\kappa_L =1$) obtained in the scaling method:
the co-oscillating (dotted lines, full), 
the boson-intrinsic (dashed lines), and
the fermion-intrinsic  (solid lines) modes.
The open squares and full circles denote the numerical results 
obtained in the TDGP+Vlasov calculation 
for the boson- and fermion-intrinsic frequencies,
which are taken from Refs. \cite{tomoBF} (monopole) 
and \cite{QPBF} (quadrupole).
In panel (c), 
the frequencies of the breathing oscillations when $\kappa_L =1/6$ 
obtained in the scaling method are shown.
}
\label{ScFr}
\end{center}
\end{wrapfigure}

From Eq.~(\ref{OsE}), which is harmonic for the variables $\vlambda$, 
the oscillation frequency $\omega$ 
and the corresponding amplitude $\vlambda$ are determined 
from the characteristic equation:
\begin{equation}
     \left( B \omega^2 - C  \right) \vlambda  = 0.
\label{eigEq}
\end{equation}
We solve the above equations numerically in the case of 
the same frequency traps for bosons and fermions ($\omega_f = 1$).
In Fig.~\ref{ScFr}, 
we show the frequencies of the monopole (a) and quadrupole
(b) oscillations in the spherically-symmetric case ($\kappa_L=1$);
the dotted lines represent the frequencies of the co-oscillating modes.
In comparison, 
we show the frequencies of the boson and fermion intrinsic modes 
when the two modes are decoupled (the dashed and solid lines, respectively).

In the quadrupole oscillation (b)
the frequencies of the co-oscillating modes (dotted line) have almost the same frequencies 
with the boson- /fermion-intrinsic modes, and show no level crossing points.

In the monopole oscillation (a), on the other hand, 
the frequencies of the boson and fermion intrinsic modes 
have crossing points at $h_{BF}/g_{BB} \approx -1.5$ and $2.2$ respectively.
In the region of $-1.5 \lesssim h_{BF}/g_{BB} \lesssim 2.2$,  
the co-oscillating collective oscillation almost agrees 
with the intrinsic modes, but, in the region,  
$h_{BF}/g_{BB} \lesssim -1.5$ and $2 \lesssim h_{BF}/g_{BB}$, 
the frequencies of the co-oscillating modes are different 
from those of the boson-/fermion-intrinsic modes.

In the same figure, 
we plot the boson- and fermion-intrinsic frequencies 
obtained in the TDGP+Vlasov approach (the open squares and 
the full circles) \cite{tomoBF,QPBF}.
We find that the scaling method well-reproduces 
the results in the TDGP+Vlasov approach 
for the BF attractive interaction, 
but gives the different values 
when $h_{BF}/g_{BB} \gtrsim 0$ (monopole oscillation) 
and $h_{BF}/g_{BB} \gtrsim 1$ (quadrupole oscillation).
The similar behavior has also been found
in the dipole oscillations \cite{tbDPL}, 
and we have explained about this discrepancy 
in the previous papers \cite{tbDPL,QPBF}.

In the last panel (Fig.~\ref{ScFr}),  
we show the frequencies when $\kappa_L=1/6$.
For convenience we refer to the four collective states as state-1 $\sim$ -4 
in high-to-low order of their oscillation frequencies;
the level crossings are found in the states-1 and -2 at 
$h_{BF}/g_{BB} \approx 0$ and $1$.

\begin{figure}
\begin{center}
{\includegraphics[scale=0.46,angle=270]{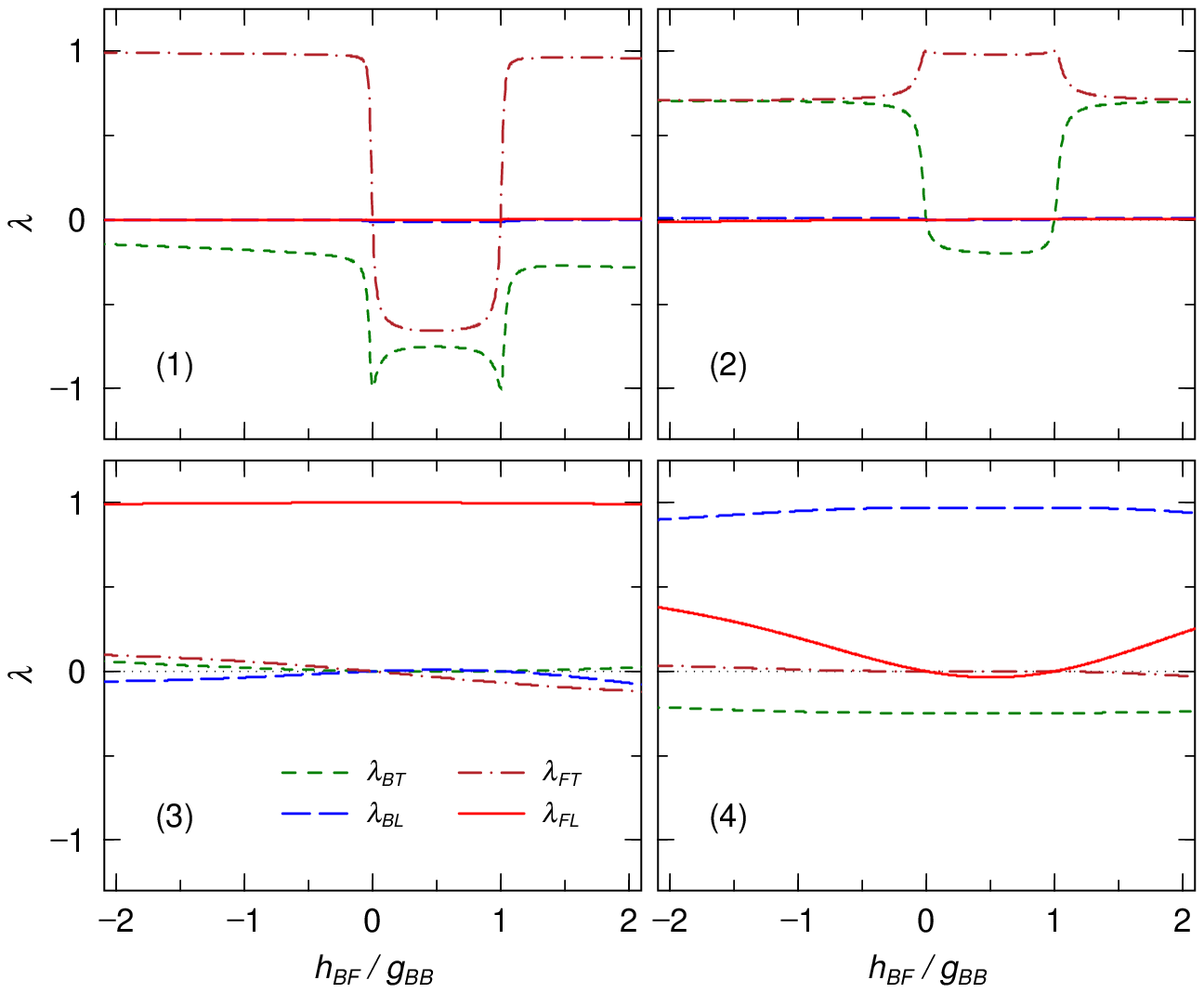}}
\caption{\small (Color online)
Components of the eigenvector of the excited states with the highest
 excited energies (1), the second highest one (2), the third highest
 one (3) and the lowest one (4) when $\kappa_L = 1/6$.  
The dashed, long-dashed, chain-dotted and solid lines 
indicate the FTB, FLB, BTB and BLB components $(\lambda_{FT}$, $\lambda_{FL}$, $\lambda_{BT}$, $\lambda_{BL})$.}
\label{ScAmpL}
\end{center}
\end{figure}

In Fig.~\ref{ScAmpL} we show the components of the eigenvector $\vlambda$
of the state-1 (1), the state-2 (2), the state-3 (3) and the state-4 (4). 
From the panels (1) and (2), 
we find that the state-1 and the state-2 are the transverse modes;
when $0 \lesssim h_{BF} / g_{BB} \lesssim 1$, the relative phases between the BTB and FTB oscillations 
are in-phase ($\lambda_{FT} \approx \lambda_{BT}$) for the state-1 
and out-of-phase ($\lambda_{FT} \approx -5 \lambda_{BT}$) for the state-2.
In the other region of $h_{BF}/g_{BB}$ 
these two states show the exchange of oscillations characters, 
which originated in this exchange of the level crossing 
of the frequencies of the state-1 and -2.
We call them the in-phase transverse breathing (ITB) and out-of-phase transverse breathing (OTB) modes.

Fig.~\ref{ScAmpL}-(3) shows that the component $\lambda_{FL}$ dominates the eigenvector, 
so that the state-3 is almost composed of the FLB mode.
The eigenvector of the state-4 shows complicates behaviors.
From Fig.~\ref{ScAmpL}-(4), 
it is found to be composed of the BLB and BTB modes,
and also include the FLB modes when $|h_{BF}|$ is large.
When $h_{BF} = 0$, the eigen-energy of the state-4 becomes  
\begin{eqnarray}
     \omega^2 =  
          \left( 1 - \frac{ v_B}{10} \right) 
          \left( 2 +\frac{3\kappa_L^2}{2} 
                     - \sqrt{4  -4 \kappa_L^2 +\frac{9 \kappa_L^4}{4} }
          \right) ,
\label{BsolSC}
\end{eqnarray}
with $v_B = V_{bb}/X_B$.
In the limit of large deformation ($\kappa_L \ll 1$),
it becomes
$\omega^2  \approx  5 \kappa_L^2  (1 -v_B/10) /2$, 
and the eigenvector of the state-4 is approximated by
\begin{equation}
     \left( \begin{array}{c} 
            \lambda_{BT} \\ 
            \lambda_{BL} 
            \end{array} \right)
     \approx \left( \begin{array}{c} 
                   -1 +\frac{10}{17} \kappa_L^2 \\ 
                    4 +\frac{5}{34}\kappa_L^2 
                    \end{array} \right) 
          \approx  \left( \begin{array}{c} -1 \\ 4 \end{array} \right),
\label{BABAmp}
\end{equation}
where
the bosons oscillates in both the longitudinal and transverse directions 
in out-of-phase. 
This mode is similar to the quadrupole oscillation, 
but the transverse-to-longitudinal amplitude ratio  ($-1 : 4$)
is different.
According to Ref.~\cite{Hu04}, 
we call it the boson axial-breathing (BAB) mode. 
When $|h_{BF}/g_{BB}| \ll 1$ and $\kappa_L \ll 1$,
the contribution of the FLB mode can be evaluated 
in perturbation method:
\begin{equation}
\lambda_{FL} \approx \frac{3 V_3}{2 \tlX_F} .
\label{VLF1}
\end{equation}
From Eqs.~(\ref{dFdx}), (\ref{V3}) and  (\ref{VLF1}), 
we find that $\lambda_{FL} < 0$ when $0<h_{BF}/g_{BB}<1$, 
and $\lambda_{FL} > 0$ otherwise. 
The obtained estimation qualitatively agrees with the numerical result 
(solid line in Fig.~\ref{ScAmpL}d).

Thus, the scaling method predict four kinds of collective oscillation modes, 
ITB, OTB, FLB and BAB modes.
In the next section we examine this description in the time-dependent approach.

\newpage

\section{Breathing Oscillations in Dynamical Approach}

\subsection{Time-Dependent Equations}
\label{TevEq}

In this section we describe the time evolution of the system 
using the TDGP equation for the boson condensate 
and the Vlasov equation for the fermions \cite{KB}:
\begin{eqnarray}
     i \frac{\partial}{\partial \tau}  \phi_c (\vbr, \tau) &=&
     \left\{ - \frac{1}{2} \nabla_r^2 + U_B (\vbr) \right\} ~ \phi_c (\vbr, \tau) ,
\label{TDGP}\\
     \frac{d}{d \tau} f(\vbr,\vp;\tau) &=&
     \left\{ \frac{\partial}{\partial \tau} + {\vp}\cdot{\nabla_r} -
     [\nabla_r U_F(\vbr)][\nabla_p] \right\} f(\vbr,\vp;\tau) = 0 ,
\label{Vlasov}
\end{eqnarray}
where $f(\vbr,\vp;\tau)$ is the fermion phase-space distribution function, 
and the $U_B$ and $U_F$ are effective potentials for bosons and fermions:
\begin{eqnarray}
     U_B (\vbr) &=& \frac{1}{2} (\vbr_T^2 +\kappa_L^2 z^2)
                 +g_{BB} \rho_B (\vbr) 
                 +h_{BF} \rho_F (\vbr) ,
\label{uB}\\
     U_F (\vbr) &=& \frac{1}{2} m_f \omega^2_f 
                    (\vbr_T^2 + \kappa_L^2 z^2)
                 +h_{BF} \rho_B (\vbr) .
\label{uF}
\end{eqnarray}

In order to solve the Vlasov equation (\ref{Vlasov}) numerically,
we use the test-particle method \cite{TP};
for each fermion, we prepare ${\tilde N}_T$ test particles per fermion  
whose coordinates and momenta are $\vbr_i$ and $\vp_i$, 
Then the fermion phase-space distribution function is described as
\begin{equation}
     f(\vbr,\vp,\tau) = \frac{(2 \pi)^3}{\Nt} 
     \sum_{i=1}^{{\tilde N}_T N_f} \delta\{\vbr-\vbr_i(\tau)\} \delta\{\vp-\vp_i(\tau)\} .
\label{TP-Wig}
\end{equation}
Substituting Eq.(\ref{TP-Wig}) into Eq.(\ref{Vlasov}),
we can obtain  the equations of motion for test-particles:
\begin{equation}
     \frac{d}{d \tau} \vbr_i (\tau) =\frac{\vp_i}{m_f}, \quad
%
     \frac{d}{d \tau} \vp_i (\tau) =- \nabla_r U_F(\vbr).
\label{eqM2}
\end{equation}

As the initial conditions of time-development at $\tau= 0$, 
we use the boosted condensed-boson wave function 
and the fermion test-particle coordinates: 
\begin{eqnarray}
      \phi_c(\vbr, \tau=0) &=& \exp\left\{ \frac{i}{2}(b_T r_T^2 + b_L z^2) \right\} 
                            \phi_c^{(g)} (\vbr), 
\label{bin}\\
     \vp_T(i) = \vp_T^{(g)}(i) + m_f \omega_f c_T \vbr_T (i), &&
     p_z (i) = p_z^{(g)}(i) + m_f \kappa_L \omega_f c_L z (i) ,
\label{fin}
\end{eqnarray}
where $b_T$, $b_L$, $c_T$ and $c_L$ are the boost parameters,
and the superscript $(g)$ represents the ground state.

In order to examine the aspects of the breathing oscillation,
we introduce the quantities:
\begin{equation}
     \Delta x_{L,T}(\tau;s) =
     \frac{1}{2} \left[ \frac{ R^2_{L,T}(\tau;s)}{R_{L,T}^{(g)2}(s)} -1 \right] 
     \approx \frac{ R_{L,T}(\tau;s)}{R_{L,T}^{(g)}(s)} -1, \quad (s=B,F)
\end{equation}
where $R_{L,T}(B)$ and $R_{L,T}$ are the root-mean-square radii of the
boson and fermion density-distributions 
along longitudinal (transverse) directions.
We should note that the last approximate term is satisfied 
when the variation is small: $|R_{L,T} - R_{L,T}^{(g)}| \ll R_{L,T}^{(g)}$.
The quantities $\Delta{x}$ represent changes of particle distributions
in longitudinal and transverse directions.

In order to inspect the oscillation modes, 
we use the strength functions defined by the Fourier transform 
of $\Delta x_{L,T}(s)$ ($s=B,F$):
\begin{eqnarray}
     S_{L,T}(\omega;s) &=& 
          \int^{t_f}_{t_i} d\tau \Delta x_{L,T}(\tau;s) 
                                 \sin{\omega \tau}
\label{stFn}
\end{eqnarray}
In actual calculation, 
we use the test-particle number ${\tilde N} = 100$ 
and take $0<\tau<200$ for the integration interval
unless otherwise noted.

In the linear-response approximation, 
only the $S_{L,T}$ have finite strengths in the present initial conditions:
$\Delta x_{L,T} = 0$ and $d \Delta x_{L,T} /d \tau \neq 0$ at $\tau=0$.

In the present paper, 
we take three kinds of mixtures: 
\Ybfst ($h_{BF}/g_{BB} = 0.573$), \Ybsnd ($h_{BF} / g_{BB} = -1.273$) and 
\Ybthd ($h_{BF} / g_{BB}= 1.325$), 
where the boson-fermion interaction is weakly
repulsive, strongly attractive and strongly repulsive, respectively;
the scattering lengths are shown in Ref.~ \cite{QPBF}.

We deal with the Yb-Yb system, 
where the number of the bosons and the fermions 
are $N_b = 10000$ and $N_f = 1000$, respectively.
The trapping potential parameters are
$\Omega_B = 2 \pi \times 300$ (Hz), $\kappa_L = 1/6$ and $\omega_f = 1$:
these values are chosen to be almost similar with those in the experiment by Kyoto group 
\cite{Fukuhara-M} (the axial symmetry is a little broken in the actual experiment).
The mass differences of the Yb-isotopes can be safely neglected, 
so that we use the same mass ($m_f=1$) for all Yb-isotopes in the present calculation.

\begin{figure}[b]
\begin{center}
{\includegraphics[scale=0.6,angle=270]{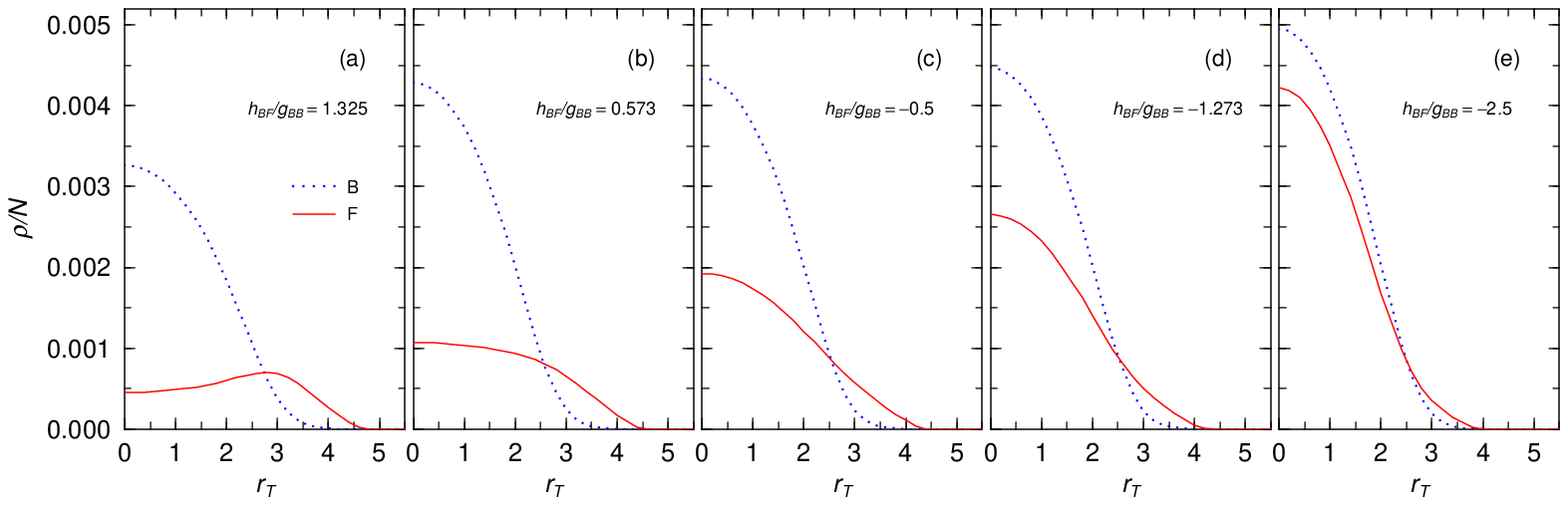}}
\caption{\small (Color online)
The ground-state density distributions of the BF mixtures
for $h \equiv h_{BF}/g_{BB} = 1.325$ (a), 
$0.573$ (b), 
$-0.5$  (c) 
$-1.273$ (d) and 
$-2.5$ (e).
The (a), (b) and (c) correspond to the $^{174}$Yb-${}^{173}$Yb, $^{170}$Yb-${}^{171}$Yb and $^{170}$Yb-${}^{173}$Yb
mixtures, respectively.
The dotted and solid lines represent are for the boson and fermion distributions.}
\label{grdRH}
\end{center}
\end{figure}

\subsection{The Ground States}
\label{GRD}

In Fig.~\ref{grdRH}, 
we show the boson (dotted line) and fermion (solid lines) density distributions 
of the ground states at $z=0$ when 
$h_{BF}/g_{BB} = 1.325$ ($^{174}$Yb$-^{173}$Yb) (a), 
$0.573$ ($^{170}$Yb$-^{171}$Yb) (b), 
$-0.5$  (c), 
$-1.273$ ($^{170}$Yb$-^{173}$Yb) (d)
and $-2.5$ (e),
where $g_{BB} = 0.154$ for Fig.~\ref{grdRH}a and 0.0964 for the other cases.
We should note that the $\omega_z z$-dependence of the density distribution at
$r_T=0$ are approximately the same as these distribution exhibited in this figure 
because of the scaling invariance discussed in the previous section.

The boson densities are center-peaked in all cases. 
In contrast, the fermion densities have a surface-peaked shape 
when $h_{BF}/g_{BB} > 1$ (Fig.~\ref{grdRH}a),
but have a center-peaked shape when  $h_{BF}/g_{BB} < 1$. 
As the BF interactions becomes attractively stronger ($h_{BF}/g_{BB} < 0$),
the fermion density is more largely distributed 
in the boson-distributed regions, and the boson density
increases at the center.
This BF interaction dependence of the fermion density
distribution can be easily explained in the TF approximation,
where the ground-state density is determined 
not by $h_{BF}$ and $g_{BB}$ independently 
but only through the ratio $h_{BF}/g_{BB}$.

In the limit of the large boson and fermion numbers, 
the density distribution is well approximated by the results in the TF method:
$\rho_{B,F} (\vbr) \approx \rho_{B,F}^{(s)} (\sqrt{\vbr_T + \kappa_T^2 z^2})$
with $\rho^{(s)}$ being that in the spherical trap ($\kappa_L=1$).
The explicit form of $\rho^{(s)}_{B,F}$ have been given in the previous paper \cite{QPBF}.

\subsection{Longitudinal Breathing Oscillations}

\begin{figure}[b]
\begin{center}
{\includegraphics[scale=0.8,angle=270]{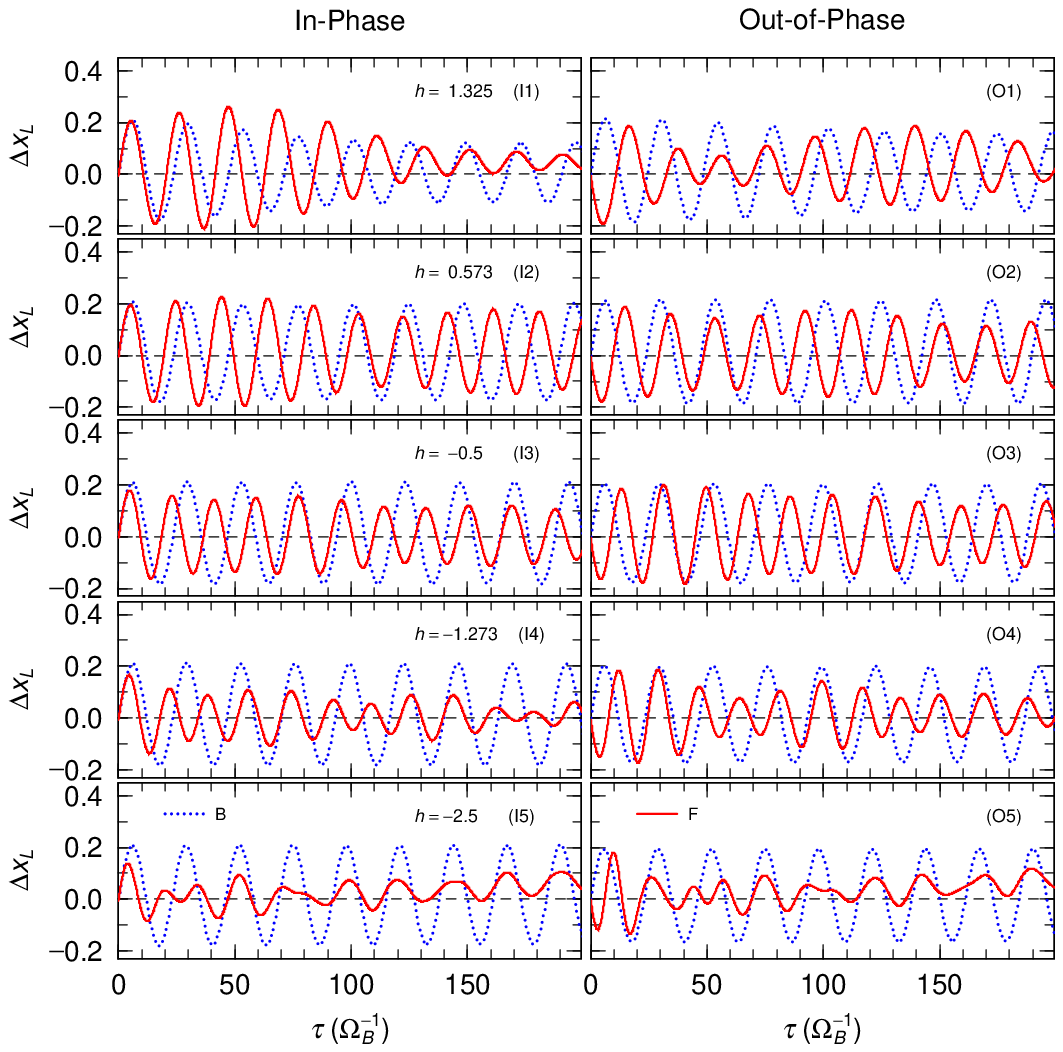}}
\caption{\small (Color online)
Time evolution of longitudinal $\Delta x_L(B)$ (boson oscillation, dotted lines)
and $\Delta x_{L}(F)$  (fermion oscillation, solid lines)
for the in-phase initial condition, 
$b_T=c_T=0$ and $b_L=c_L=0.01$, 
(left figures) and
for the out-of-phase initial condition, 
$b_T=c_T=0$ and $b_L=-c_L=0.01$, 
(right figures).
The panels from the upper to the lower columns are for 
$h \equiv  h_{BF}/g_{BB} = 1.325$ (I1 \& O1), $0.573$ (I2 \& O2), 
$-0.5$ (I3 \& O3), $-1.273$ (I4 \& O4) and $-2.5$ (I5 \& O5), respectively.}
\label{brBFL}
\end{center}
\end{figure}

First, we present numerical calculations in the TDGP+Vlasov approach 
with the longitudinally-deformed BF in-phase condition,
$b_T=c_T=0$  and $b_L=c_L=0.01$.
We note that this condition is consistent with the actual experiments 
by Kyoto group \cite{Fukuhara-M}.
For comparison, we also present the  BF out-of-phase condition,
$b_T=c_T=0$  and $b_L=c_L=0.01$.

In Fig.~\ref{brBFL}, we show the time-dependence of $\Delta x_{L}$ 
for $h_{BF}/g_{BB} = 1.325$, $0.573$, 
$-0.5$, 
$-1.273$ and $-2.5$.
The left and right figures exhibit the results for the in-phase
and out-of-phase initial conditions.
The $\Delta x_L(B)$ is found to oscillate monotonously (dotted lines),
while the $\Delta x_L(F)$ shows a small damping and/or a small beat.  
Furthermore, we can find that the $x_L(F)$ has a tendency to synchronize
with $x_L(B)$ in the case of the large BF couplings ($h_{BF}/g_{BB} \ge 1.0$),
and that the $x_L(F)$ gradually increases in $\tau \gtrsim 150$ 
when $h_{BF}/g_{BB} \ge 2.5$.

\begin{figure}[h]
\begin{center}
{\includegraphics[scale=0.8,angle=270]{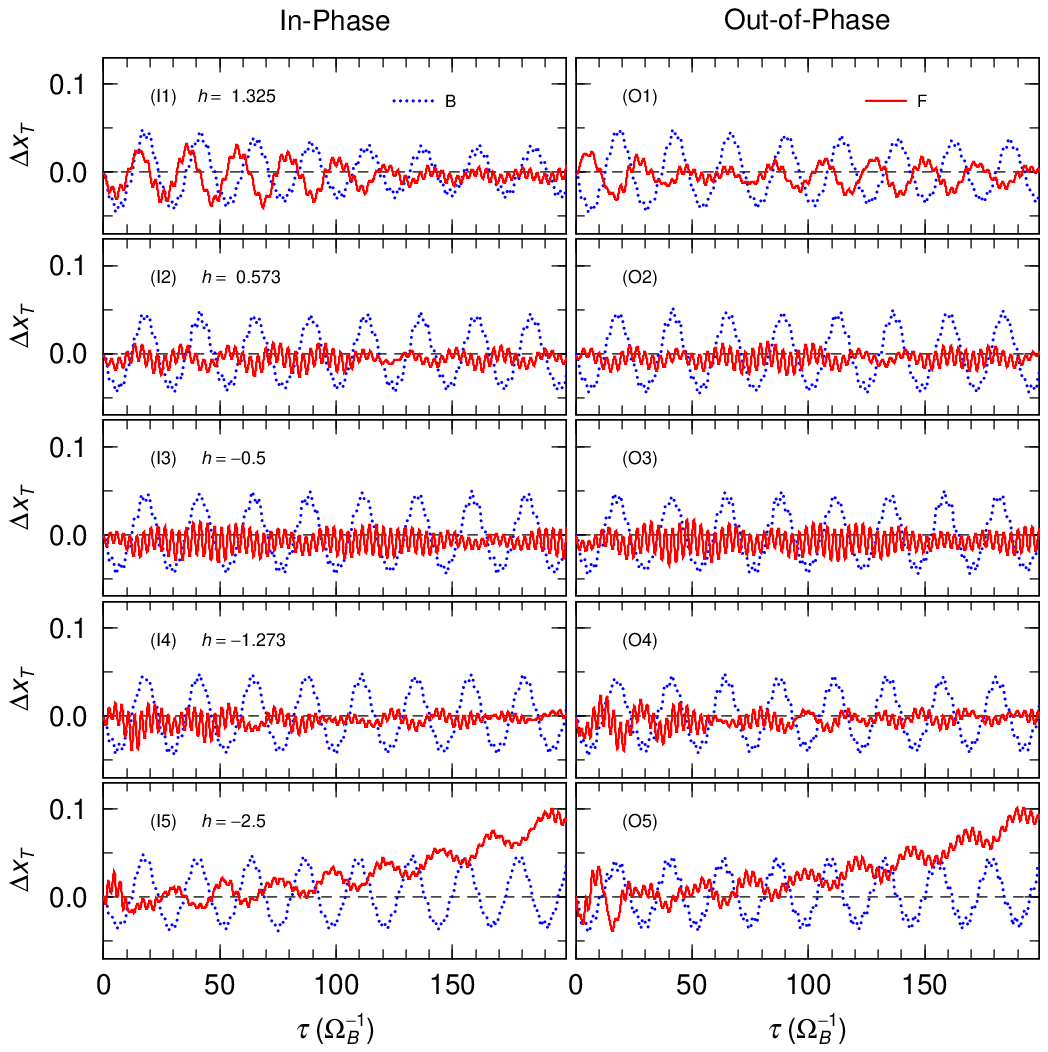}}
\caption{\small (Color online)
Time evolution of transversal $\Delta x_{T}(B)$ (boson oscillation, dotted lines)
and $\Delta x_{T}(F)$  (fermion oscillation, solid lines)
in the longitudinal oscillations.
The other conditions are the same as those in Fig.~\ref{brBFL}.}
\label{brBFLT}
\end{center}
\end{figure}

\begin{figure}[h]
\begin{center}
\includegraphics[scale=0.65,angle=270]{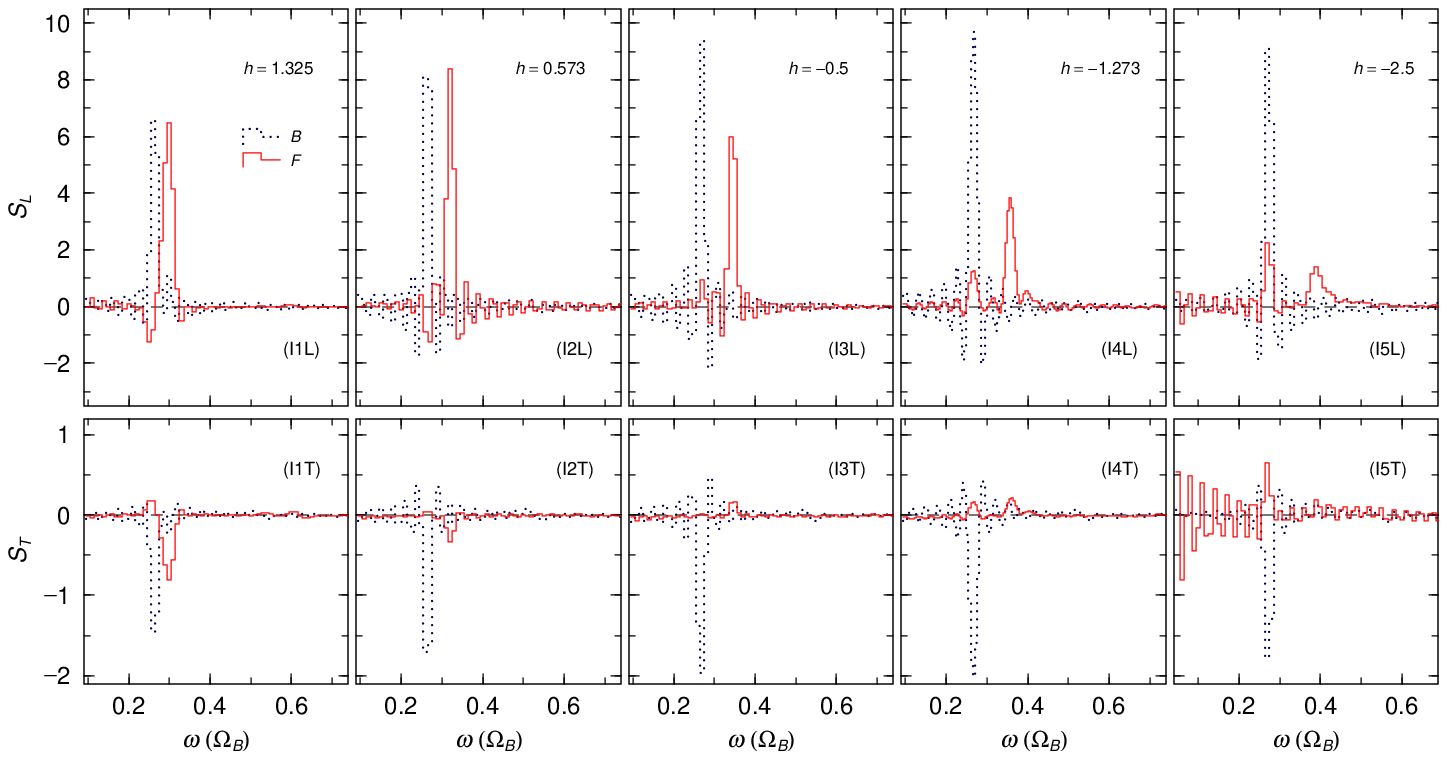}
\caption
{\small  (Color online)
The longitudinal (upper figures) and transverse strength functions
 (lower figures) of boson $S_{L,T} (B)$ (dotted lines)
and fermion  $S_{L,T} (F)$ (solid lines) (solid lines)
for the in-phase initial conditions.
The result of each panel correspond to the oscillation shown in 
the left figures of Fig.~\ref{brBFL}
 }
\label{StLFin}
\end{center}
\end{figure}

In Fig.~\ref{brBFLT}, we show the time-dependence of $\Delta x_{T}$ 
in the same oscillations with Fig.~\ref{brBFL}.
The boson $\Delta x_T(B)$  (dotted lines) monotonously oscillate 
with the same period with $\Delta x_T(F)$, but out of phase.
The fermion  $\Delta x_T(F)$ seems to be a superposition 
of two modes with different periods (dotted lines), 
which makes a beat.
The long-period mode is dominant when $|h_{BF}/g_{BB}| \gtrsim 1$,
while the short-period mode becomes dominant when $|h_{BF}/g_{BB}| \lesssim 1$. 
Furthermore, in the case of the strongly attractive BF interaction 
($h_{BF}/g_{BB}=-2.5$,I5 \& O5),
monotonous increases of $x_T(B)$ are confirmed in $\tau \gtrsim 100$, 
which means the fermion-gas expansions in the transverse direction.

\begin{figure}[h]
\begin{center}
\includegraphics[scale=0.65,angle=270]{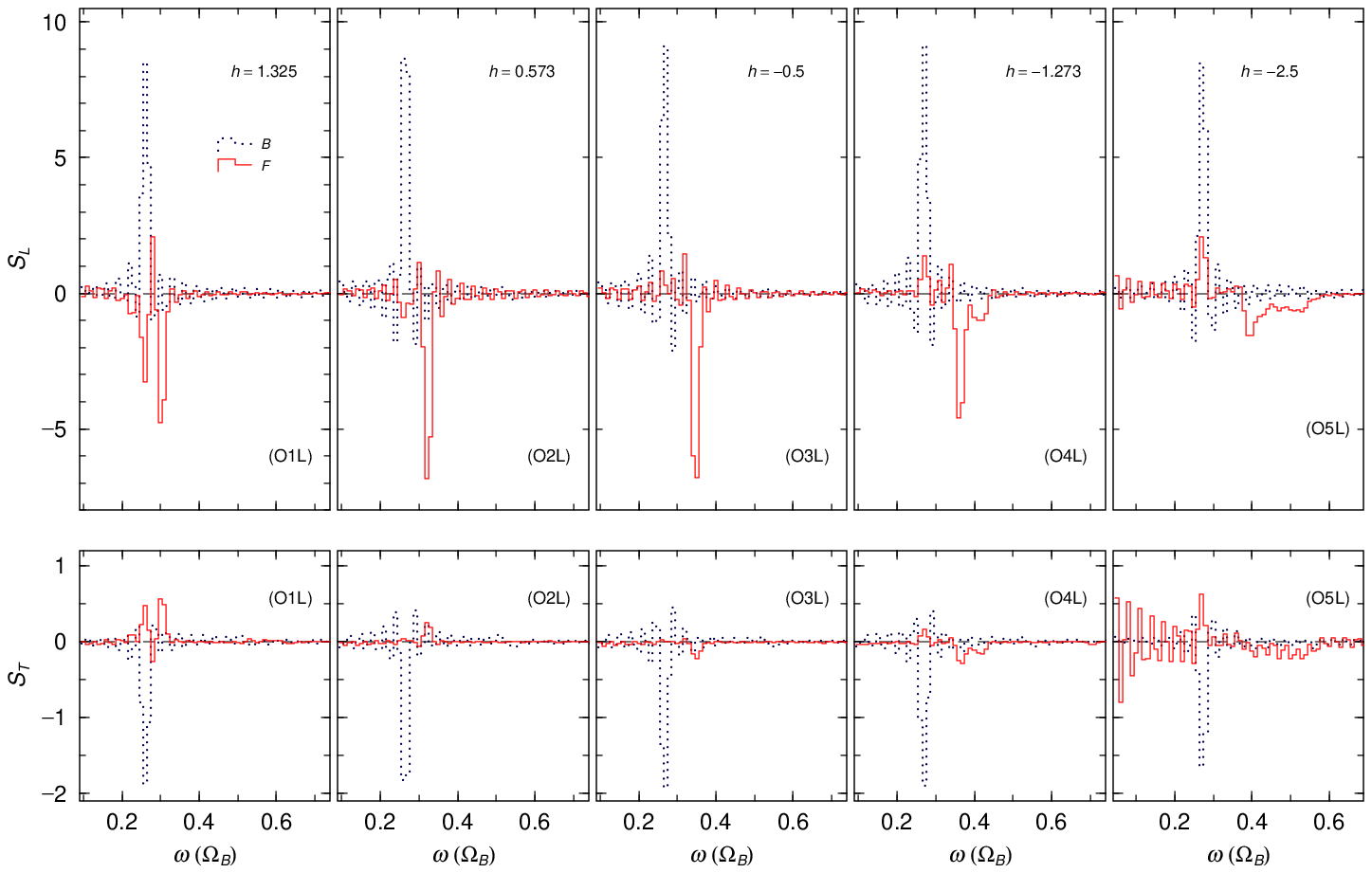}
\caption
{\small  (Color online)
Same as Fig.~\ref{StLFin}, but the initial condition is out-of-phase.
The longitudinal (upper figures) and transverse strength functions
 (lower figures) of boson $S_{L,T} (B)$ (dotted lines)
and fermion  $S_{L,T} (F)$ (solid lines) (solid lines)
for the in-phase initial conditions.
The result of each panel correspond to the oscillation shown in 
the right figures of Fig.~\ref{brBFL}. }
\label{StLFou}
\end{center}
\end{figure}

We show the longitudinal (upper panels) and transverse 
strength functions (lower panels) of the longitudinal oscillations 
for the in-phase (Fig.~\ref{StLFin}) and 
the out-of-phase (Fig.~\ref{StLFou}) initial conditions
in $h_{BF}/g_{BB} = 1.325$ (I1), $0.573$ (I2), $-0.5$ (I3), 
$-1.273$ (I4) and $-2.5$ (I5).

First, we find that the $S_L(B)$ (dotted lines) and $S_T(B)$ (dashed lines) 
have one large peak at almost the same frequency $\omega \approx 0.26 - 0.27$ 
in all mixtures.
The peak height ratios are 
$-S_L(B) / S_T(B) \approx 4.7$ (I1 \& O1), $5.33$ (I2 \& O2), 
4.8 (I3 \& O3), 4.8 (I4 \& O4), 5.31 (I5 \& O5), 
and almost independent of the initial condition.
In addition, these ratios approximately agree 
with the result obtained in the scaling method for the BAB mode, 
$\lambda_{BL}/\lambda_{BT} \approx - 4$. 

On the other hand, the $S_L(F)$ show two peaks;
one peak appears at the same frequency with the peak of $S_L(B)$, 
and the strengths $S_L(F)$ at the first peaks are positive/negative 
for the attractive/repulsive BF interaction.
Clearly the mode corresponding to the first peak is the forced oscillation 
caused by the boson oscillation 
as shown in the dipole \cite{tbDPL} and quadrupole \cite{QPBF} oscillations.
The similar behavior is appeared in the $\lambda_{FL}$
of the BAB mode in the scaling method in Eq.~(\ref{VLF1}).

The strengths at the second peak become positive/negative 
for the in-phase/out-of-phase initial conditions, 
and the boson oscillations show no strengths there.
In addition, we should note that $S_T(F)/S_L(F) < 0$ when $h_{BF}/g_{BB} >0$
and  $S_T(F)/S_L(F) > 0$ when $h_{BF}/g_{BB} < 0$;
these behaviors are the same as those of $\lambda_{FT}/\lambda_{FL}$ in the FLB
mode in the scaling method (see Fig.~\ref{ScAmpL}-(3)). 
So this mode is the intrinsic fermion oscillation mode 
corresponding to the FLB mode predicted in the scaling method.
 
Thus, the qualitative behavior of the longitudinal oscillation
almost agrees with the collective modes 
predicted in the scaling method except the sign of $S_L(F)$ in the BAB modes;
in the scaling method $\lambda_{BL}$ and $\lambda_{FL}$ have 
the same sign when $h_{BF}/g_{BB} \gtrsim 1.0$.

The other difference from the results in the scaling method is 
the fermion gas expansions mainly in the transverse directions
in the case of strongly attractive BF-interaction, 
which we discuss in the next subsection.  

\subsection{Transverse Breathing  Oscillations}

In this subsection, we discuss the oscillations caused 
by the transversely-deformed initial conditions,  
$b_{T}=c_T=0.1$ and $b_{L}=c_L=0$ (BF-in-phase condition), 
and $b_{T}=-c_T=0.1$ and $b_{L}=c_L=0$ (BF-out-of-phase condition).

\begin{figure}
\begin{center}
{\includegraphics[scale=0.8,angle=270]{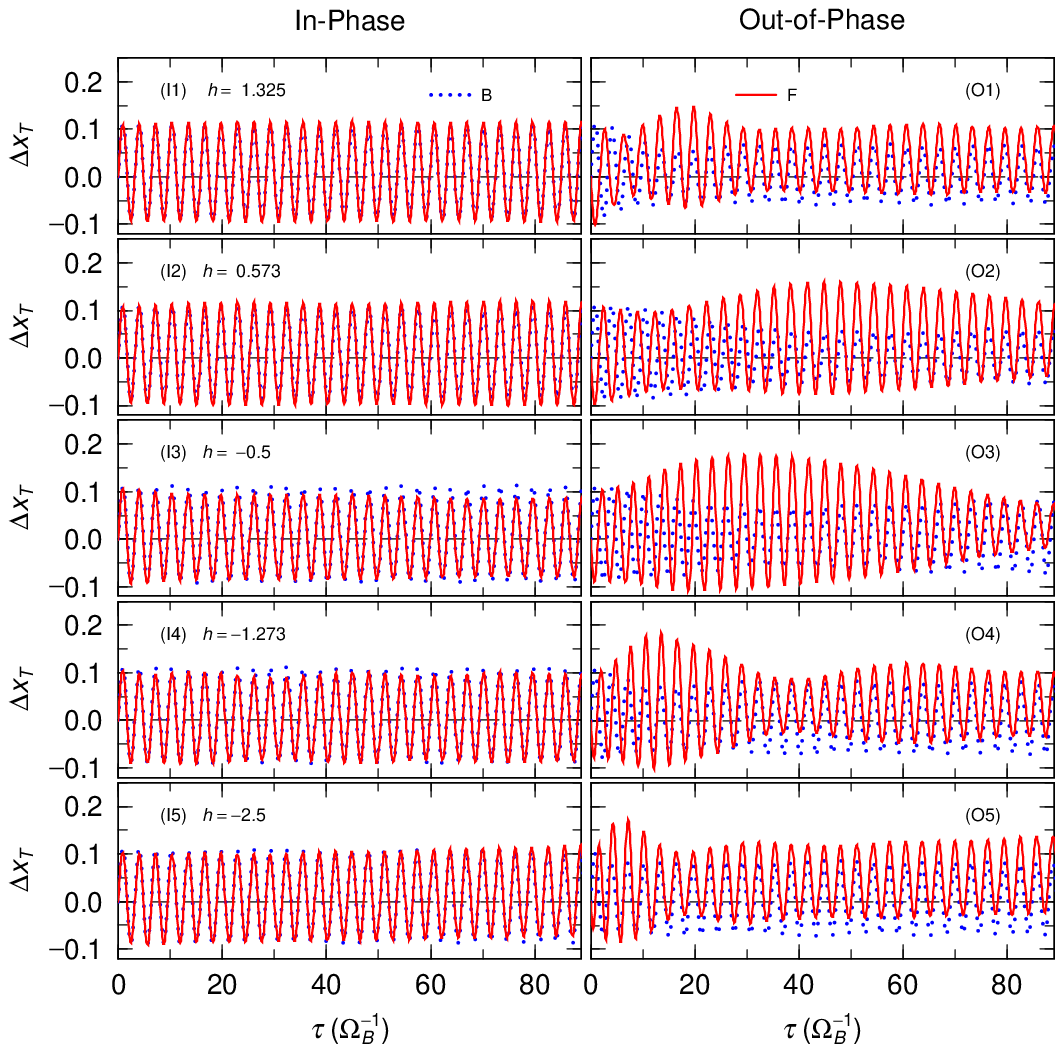}}
\caption{\small (Color online)
Time evolution of transversal $\Delta x_T(B)$ (boson oscillation, dotted lines)
and $\Delta x_L(F)$  (fermion oscillation, solid lines)
for the in-phase initial condition, $b_T=c_T=0.1$ and $b_L=c_L=0$, 
(left figures) and.  
for the out-of-phase initial condition, $b_T=-c_T=0.1$ and $b_L=c_L=0$ 
(right figures).
The panels from the upper to the lower columns are 
for $h \equiv h_{BF}/g_{BB} = 1.325$ (I1 \& O1), 
$0.573$ (I2 \& O2), $-0.5$  (I3 \& O3) 
$-1.273$ (I4 \& O4) and $-2.5$ (I5 \& O5), respectively.}
\label{brBFTs}
\end{center}
\end{figure}

In Fig.~\ref{brBFTs}, we show the time-dependence of $\Delta x_T$ 
for $h_{BF}/g_{BB} = 1.325$, $0.573$, $-0.5$, $-1.273$ and $-2.5$.
The left and right figures exhibit the results with the in-phase
and out-of-phase initial conditions, respectively.
The $\Delta x_T(B)$ is found to oscillate monotonously (dotted lines),
while the $\Delta x_T(F)$ shows a small damping and/or a small beat.  

These oscillations also excite the longitudinal-oscillation modes of 
very small amplitudes. 
They are not synchronized with the transverse oscillations, 
and their qualitative behavior 
is the same as the oscillation modes discussed in the previous section.

We find that
the oscillations with the in-phase initial condition is monotonous, 
and those with the out-of-phase initial condition
show a beat in early stage of time and become monotonous in the latter stage of time,
where relative phases between boson and fermion oscillations become in-phase.
In addition, the positive shifts of the mean values appear in $\Delta x_T$ 
for the out-of-phase initial conditions.

\begin{wrapfigure}{r}{8.6cm}
\vspace*{-0.5cm}
\begin{center}
{\includegraphics[scale=0.6]{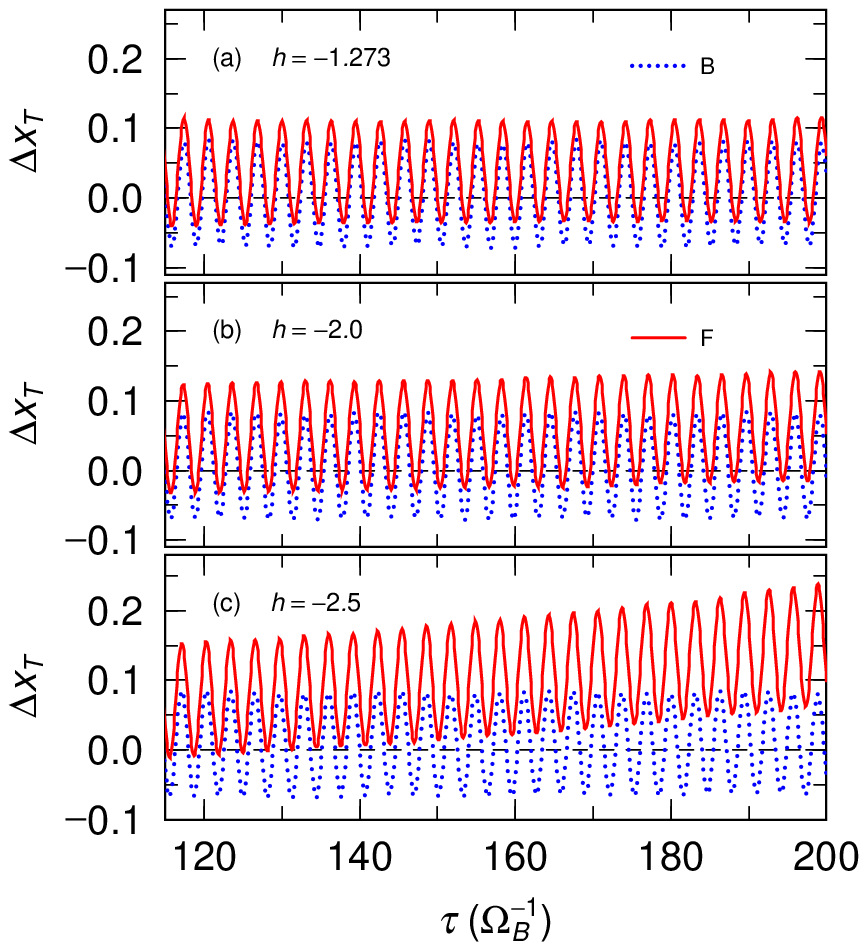}}
\caption{\small (Color online)
Time evolution of $\Delta x_{T}(B)$ (boson oscillation, dotted lines)
and $\Delta x_{T}(F)$  (fermion oscillation, solid lines)
for the in-phase initial condition.
The panel (a), (b) and (c) are for $h \equiv h_{BF}/g_{BB} = -1.273$,  
$-2.0$ and $-2.5$.}
\label{brBFTd}
\end{center}
\end{wrapfigure}

In Fig.~\ref{brBFTd} we show the same oscillations in the later time, 
$\tau \ge 115$.
We can confirm clearly the relative in-phases behaviors between 
the boson and fermion oscillations 
and the positive shift of the mean values of $\Delta x_T$.
Furthermore, we find the expansion of the fermion gases for the transverse direction, 
which has been shown also in the longitudinal oscillations.
In Fig.~\ref{brBFTd} the fermion gas expansions appear moderately 
for $h_{BF}/g_{BB} = -2$  and remarkably in the BF mixtures 
of the strongly attractive BF interactions ($h_{BF}/g_{BB} = -2.5$) 
but, in the case of the weak interactions ( $h_{BF}/g_{BB} = -1.273$), 
no expansions are observed at least in the time intervals of $\tau \ge 115$.

\begin{figure}[t]
\begin{center}
{\includegraphics[scale=0.7,angle=270]{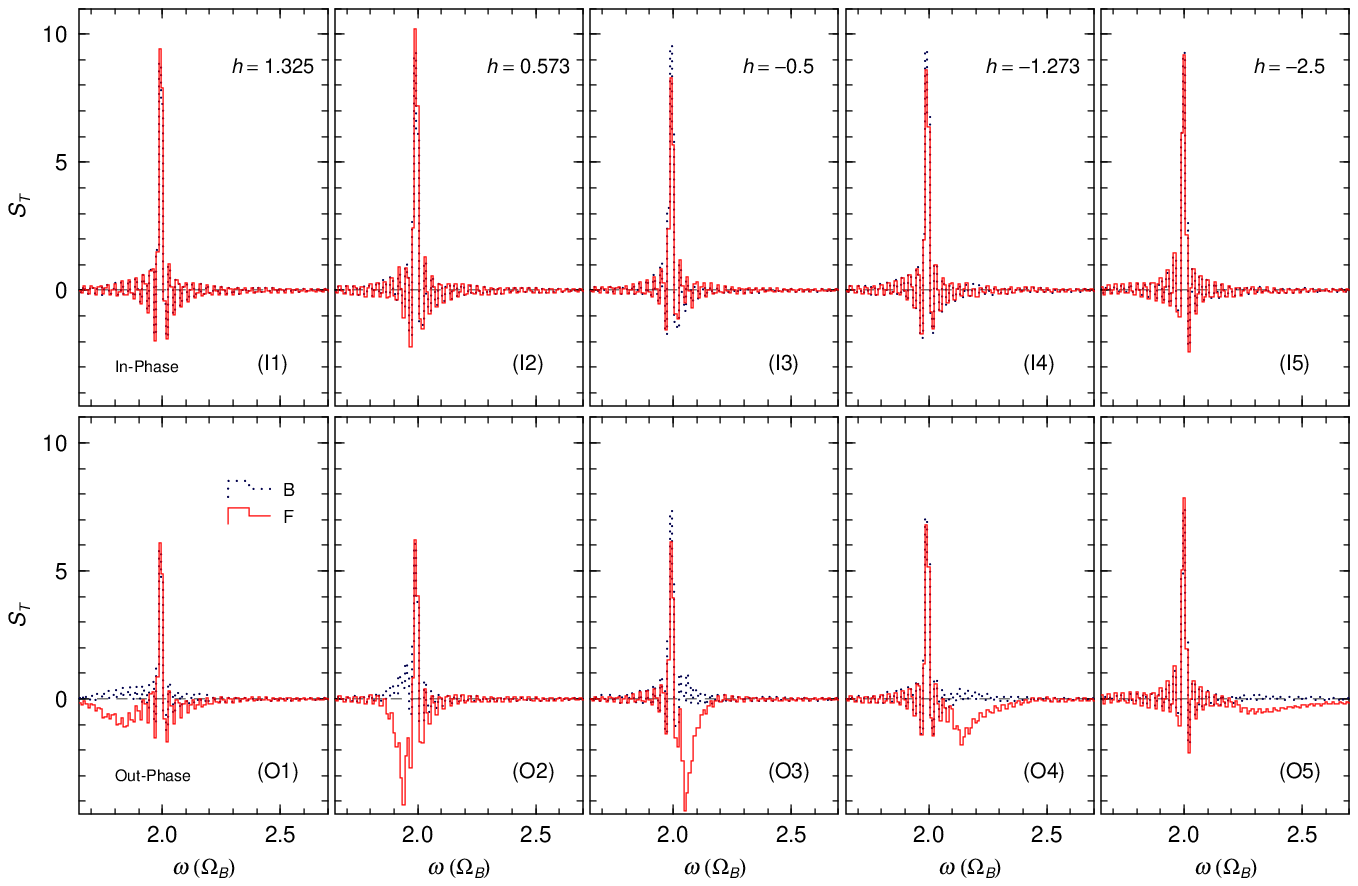}}
\caption{\small (Color online)
Strength Functions of the transverse strength functions for the boson
 (dotted lines) and fermion oscillations (solid lines)
for the in-phase initial condition (upper figures) 
and for the out-of-phase initial conditions (lower figures)
with $h \equiv h_{BF}/g_{BB} = 1.325$ (I1 \& O1), $=0.573$ (I2 \& O2), 
$=-0.5$ (I3 \& O3), $= -1.273$  (I4 \& O4) and $=-2.5$ (I5 \& O5).}
\label{StBFT}
\end{center}
\end{figure}

In order to examine the oscillation behaviors, 
we show the transverse strength functions $S_T$ in Fig~\ref{StBFT}. 
In all cases of the in-phase and out-of-phase initial conditions 
and of the BF interactions, 
the boson and fermion strength functions,
$S_T(B)$ and $S_T(F)$, have one sharp peak 
at the same frequencies, 
and $S_T(B) \approx S_T(F)$ at the peaks.
Furthermore, a broad peak exists in every oscillation
for the out-of-phase initial condition; 
the $S_L(B)$ and $S_T(B)$ take opposite values in the peak area
and the width becomes large as $|h_{BF}/g_{BB}|$ increases.

The present behaviors of the transverse oscillations can be explained
from the superposition of the ITB and OTB oscillation modes;
the strengths of the ITB modes concentrates on the narrow energy region,
and $S_T(B)$ and $S_T(F)$ have the almost same values,
while the strengths of the OTB modes distribute in broad energy region, 
and is dominant in the fermion strength function.
The ratios between $S_T(B)$ and $S_T(F)$
are qualitatively similar to those of $\lambda_{BT}$ to 
$\lambda_{FT}$ obtained in the scaling method.

The narrow ITB and broad OTB peaks explain the behaviors of $\Delta x_T(B)$ 
and $\Delta x_T(F)$ observed in Figs.~\ref{StBFT}.
When the relative phase between the boson and fermion oscillations is in-phase 
at the initial time, 
the oscillation starts only with the ITB mode, which is monotonous.
But, for the out-of-phase initial conditions, 
the oscillation includes the ITB and OTB modes at the beginning, 
which causes a beat phenomenon in the $\Delta x_T(F)$ in early stage of time.
Then, the OTB modes damps rapidly, 
the transverse oscillation becomes monotonous and relatively in-phase 
between the boson and fermion oscillations.   

\begin{wrapfigure}{r}{7.5cm}
\vspace*{-0.5cm}
\begin{center}
{\includegraphics[scale=0.4]{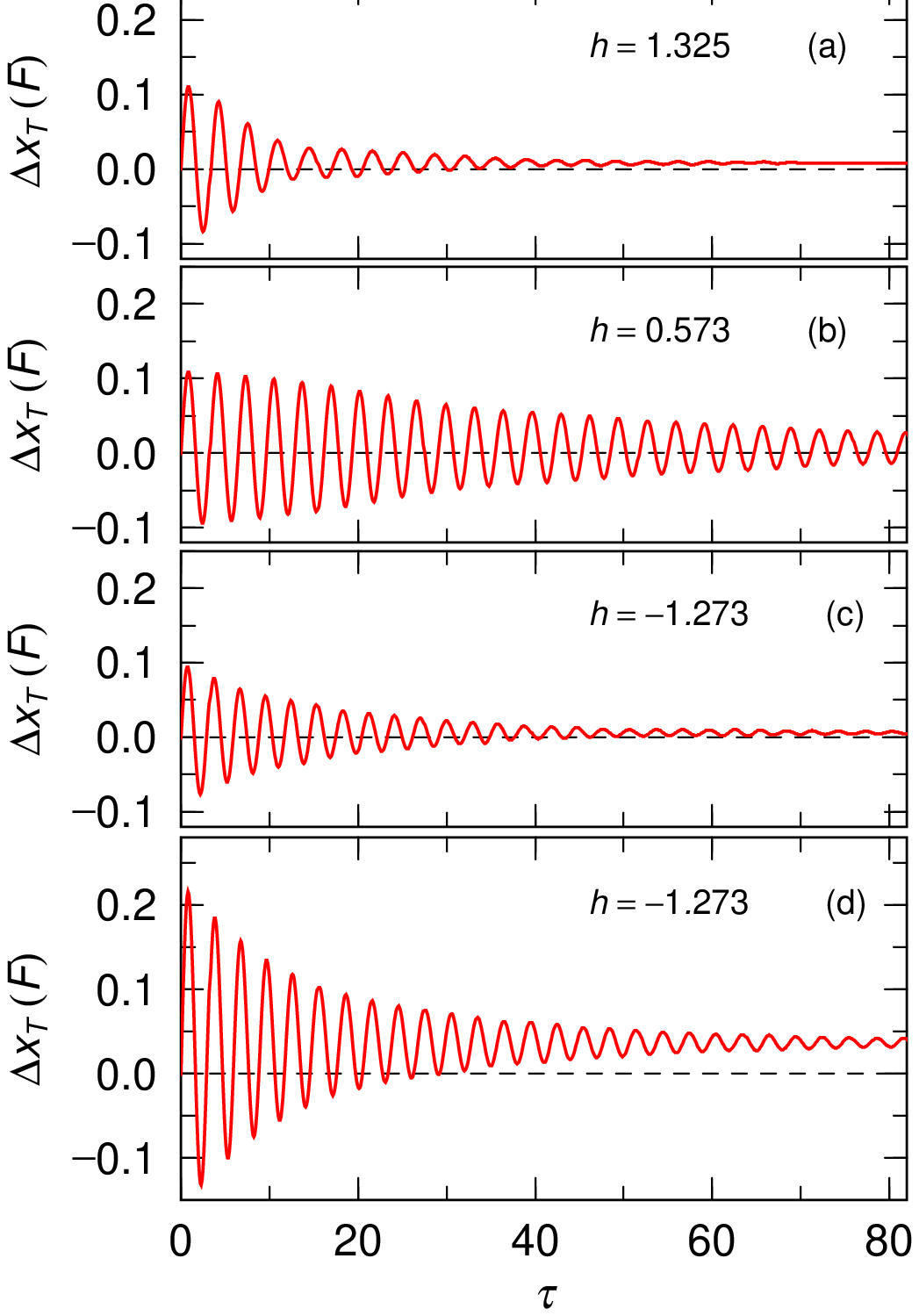}}
\caption{\small (Color online)
Time evolution of $\Delta x_{T}(F)$ in the case of frozen boson motions 
with the transversal initial conditions: $c_L=0$ and $c_T=0.1$ for the panel (a-c)
and $c_L=0$ and $c_T =0.2$ for the panel (d).
The BF interactions are $h \equiv h_{BF}/g_{BB} = 1.325$ (a), $0.573$ (b), 
$-1.273$ (c,d). }
\label{brTFS}
\end{center}
\end{wrapfigure}

As shown before, 
the OTB mode has very small boson strength, 
so that we can consider it as a fermion intrinsic mode.
In Fig.~\ref{brTFS}, 
we show the time evolution of $\Delta x_T(F)$ 
when the boson motion is frozen.
The dampings are more rapid for $h_{BF}/g_{BB} = 1.325$ (a) and 
$-1.273$ (c) than that for $h_{BF}/g_{BB} = 0.573$ (b).
The effective potentials for fermion are close to the harmonic
oscillator one in weak BF interactions and 
gives very slow dampings of the fermion oscillation, 
but the potentials for the strong BF interaction causes rapid damping effect 
through their large anharmonicities. 
In the bottom panel Fig.~\ref{brTFS}-(d), 
the calculation is shown for $h_{BF}/g_{BB} = -1.273$ 
with the initial amplitude twice larger than in panel (c);
we find the shift of the equilibrium positions of oscillation 
in the outward direction larger than in panel (c).
It shows that the slight shift of the equilibrium positions of oscillation 
in $\Delta x_T$ in the case of the out-of-phase initial condition is caused
by the energy transfer from boos gas to fermions 
through the process of the damping of the OTB mode.

In our previous studies of the BF mixtures in the spherical traps \cite{tbDPL,QPBF},
the oscillations can be described by the superposition of the boson and fermion intrinsic modes.
In the present calculation in the time-dependent approach, 
we find a phenomenon that these intrinsic modes are coupled 
and appear as the in-phase and out-of-phase intrinsic modes.
In the quadrupole oscillations with the spherical trap,
furthermore, the relative phase between the boson and fermion
oscillations becomes in-phase or out-of-phase at the later stage of time
corresponding to the attractive or repulsive boson-fermion interactions \cite{QPBF};
in contrast, the relative phase in the monopole oscillations are in-phase 
independently of the BF interactions \cite{MSY1}.

In the spherical trap,
the region of increasing boson density are close to that 
of decreasing density in the dipole and quadrupole oscillations.
In these cases, 
the fermions easily move from the dense region 
to the dilute region of the boson density 
when the BF-interaction is repulsive.
In the monopole oscillation, 
on the other hand, 
change of the boson density occurs uniformly.
In fact the relative phase between the boson and fermion oscillations
is not definite and depends 
on the strength of the BF interaction and the boson/fermion numbers et. al.

In the largely prolate deformed trap, 
expansions and shrinks of the fermion density occurs easily 
in the transverse direction 
in the case of the longitudinal breathing oscillations,
but such an anisotropy of the fermion flow is difficult 
in the transverse breathing oscillations.
Thus, in the largely deformed trap, 
the longitudinal and transverse breathing oscillations 
show the similar behaviors 
with the quadrupole and monopole ones in the spherical trap, respectively.

Furthermore, the fermion-gas expansion 
in the case of strongly attractive BF-interactions has also appeared 
in the quadrupole oscillation in the spherical trap \cite{QPBF} 
though this mode is time-even.
In the spherical trap, 
this expansion is a part of the monopole oscillation, 
which is decoupled from the quadrupole oscillation and 
not included at the beginning. 
In contrast, in the largely deformed trap, 
the expansion mode is a part of the FLB oscillation, 
which is coupled with the other modes, 
so that this expansion appears also in time-odd modes. 
Indeed, the expansion mode appears in the monopole oscillation with 
the time-even initial condition in the spherical trap 
when the initial condition is taken to be time-odd.

\subsection{Comparison with the Scaling Method}

In this subsection, 
we discuss the BF-interaction dependence 
of the intrinsic-mode frequencies 
obtained in the TDGP+Vlasov approach, 
and compare them with those in the scaling method.

\begin{figure}
{\includegraphics[scale=0.55]{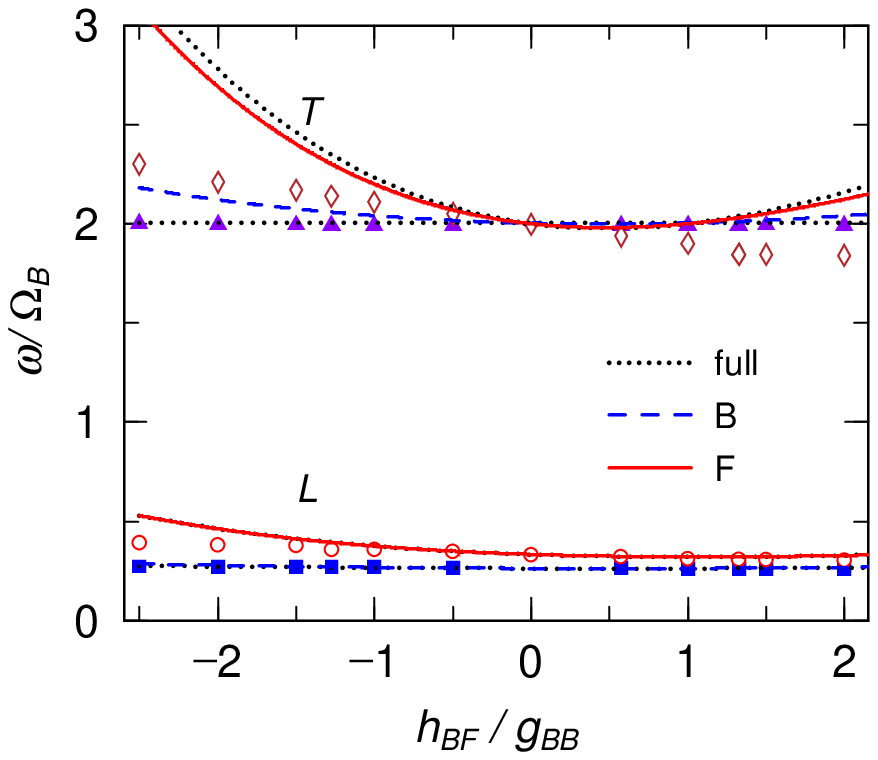}}
\caption{\small (Color online)
Intrinsic frequencies of the breathing oscillations of the boson and fermion gases
obtained in the TDGP+Vlasov approach;
full squares, open circles, solid triangle and open diamonds  are 
for the BAB, FLB, ITB and OTB modes in the 
mixtures, respectively. 
Solid, dashed and dotted lines represent the results in the scaling
 method (same as Fig.~\ref{ScFr}c).
}
\label{frCdp} 
\end{figure}

As discussed in Sec.~II, 
the breathing oscillations of the BF mixtures are described 
as the superposition of four collective modes: 
BAB (boson axial-breathing), 
FLB (fermion longitudinal breathing), 
ITB (in-phase transverse breathing) and OTB (out-of--phase transverse breathing),
which are characterized in the scaling method.  
Fig.~\ref{frCdp} shows the $h_{BF}/g_{BB}$-dependences 
of the boson and fermion intrinsic frequencies, 
which are obtained from the peak positions of the strength functions.
 
The excitation energies of the BAB and ITB modes in TDGP+Vlasov agree
with those of the state-1 and state-3 in the scaling method, respectively.
As discussed in the previous subsections, 
the mixing ratios of the boson and fermion intrinsic modes in these states are also qualitatively same in two approach,
but the excitation energies of the FLB and OTB (FTB) modes dose not agree.

The peaks of the strength functions corresponding to the BAB and ITB modes 
are narrow, and their strengths concentrate into one mode.
In contrast, the OTB peak is very broad in any condition, 
and the corresponding oscillation damps very rapidly.
In addition, the peak correspond to the FLB mode is also broad 
when $h_{BF}/g_{BB} \lesssim -1.0$.
Thus, the strength of the fermion oscillation modes do not concentrate to
one mode with the fixed frequency, and this oscillation behavior is not
applicable in the scaling method based on the sum-rule approach.

Furthermore, when the boson motion is frozen, 
the intrinsic frequency of the FLB oscillation in the scaling method  
is written as
\begin{equation}
     \omega_{FLB}^2 = 
     \frac{\int d^3 r \rho_F(\vbr) \left( 3 z^2 \frac{\partial^2 U_F}{\partial z^2} 
                      + 5 \frac{\partial U_F}{\partial z}
               \right)}{  
          2 m_f \int d^3 r z^2  \rho_F(\vbr)},
\end{equation}
where $U_F$ is the fermion potential defined as Eq.~(\ref{uF}).
Thus, the oscillation frequency is reflected by the first and second
derivative of fermion potentials.

In the scaling method, the amplitude of the oscillation is assumed to be
very small, and the $\omega_{FLB}$ is reflected by 
the potential shape in the region where the fermions occupy.
When the amplitude becomes larger, however, the $\omega_L$ is influenced
from the potential shape outside the fermion occupation region, and
its value approaches to $2 \kappa_L \omega_f = 0.333$.
 The FTB oscillation also shows similar behavior, and its frequency approaches
$2 \omega_f = 2$ as the amplitude becomes larger.

The frequencies of the fermion intrinsic oscillation modes monotonously 
decrease as the $h_{BF}/g_{BB}$ becomes larger 
in the TDGP+Vlasov approach
though the frequencies decrease when $h_{BF}/g_{BB} \lesssim 0.5$ and
increase $h_{BF}/g_{BB} \lesssim 0.5$ in the scaling method. 
This behavior of the fermion frequencies comes from the structure change  
of the fermion ground-state density distribution 
into the surface-peaked shape (Fig.~\ref{grdRH}). 

In the scaling method calculation, 
the frequency of the FLB and FTB (OTB) mode has the minimum around 
$h_{BF}/g_{BB} \sim 0.5$, 
and increases with increase of $h_{BF}/g_{BB}$ 
in the region of $h_{BF}/g_{BB} > 1$ (Fig.~\ref{frCdp}, solid line);
On the other hand, in the TDGP+Vlasov approach, 
the density distribution and the velocity field varies in time 
in large amplitude oscillations.
In the mixture of $h_{BF}/g_{BB} > 1$, 
the minimum position of the fermion potential
is on the spherical surface at the border of the boson density distribution,
and the fermion gas oscillates around this surface; 
it results in the frequencies of the fermion modes 
smaller than those obtained in the sum-rule.
The same discrepancies appears in the dipole and quadrupole oscillations
in the spherical trap\cite{tomoBF, monoEX, tbDPL, QPBF}.

Finally, we comment on the experiment by Fukuhara et.al. in Ref.~\cite{Fukuhara3},
where the frequencies of the BLB oscillations was obtained.
The experimental results show
that $\omega_{BAB} = 1.69 \kappa_L \Omega_B$ when $h_{BF}/g_{BB} = 0$ and
 $1.67 \kappa_L \Omega_B$ when  $h_{BF}/g_{BB} = 1.325$ (\Ybthd),
while the present calculations are $\omega_{BAB} = 1.59 \kappa_L \Omega_B$ for $h_{BF}/g_{BB} = 0$ 
and $1.57 \kappa_L \Omega_B$ for $h_{BF}/g_{BB} = 1.325$.
The experiments have been performed using the non-axial symmetric trap 
with the different atomic numbers at finite temperature.
Then, we cannot make direct comparison between the two results, 
but we can mention that the two results qualitatively agree each other, 
particularly in the 1.2\% increase when the BF coupling is varied from 
$h_{BF}/g_{BB} = 0$ to  $h_{BF}/g_{BB} = 1.325$.

\section{Summary}

In this paper, 
we investigated the collective breathing oscillation 
of the BF mixtures of Yb isotopes in the deformed trap ($\kappa_L=1/6$)
by varying the BF-coupling constant.

The scaling method predicts that the couplings between BLB and BTB 
modes and between the BTB and FTB modes make four intrinsic modes, 
BAB, FLB, ITB and OTB modes.
This prediction is qualitatively consistent with the results in 
the TDGP+Vlasov approach.

In the case of the weak boson-fermion interactions, 
the longitudinal and transverse breathing oscillations decouple 
in the largely deformed trap of prolate shape.
Because of the large interaction energy coming from the condensed bosons, 
the boson breathing oscillations couple into the intrinsic BTB and BAB modes, 
and the intrinsic FTB and FLB modes are made from the fermion oscillations.  
The actual time-development processes are described 
by the superposition of these intrinsic modes.

The intrinsic transverse modes (BTB \& FTB) have very close frequencies, 
and it is similar in the longitudinal modes (BAB \& FLB);
the frequencies of these two pairs separate very well. 

In the spherical trap,
the fermion oscillation have been shown to have
the boson-forced oscillation modes 
and the two intrinsic modes 
which correspond to the inside- and outside-fermion oscillations 
for the boson-distributed regions.
In the present prolate deformed trap, 
no outside-fermion modes appear.

In the longitudinal breathing oscillations  a beat and a small damping appear
for both the boson an fermion oscillations though the boson oscillations 
shows these behaviors much weaker than the fermion oscillation.
The intrinsic frequencies of the BAB and FLB modes are very close, 
and the influence from the boson
oscillation to the fermion oscillation is large while the opposite
influence is weak.
The FLB oscillation includes the two mode with the close frequencies and
show the above behaviors.
 
The transverse breathing oscillations show monotonous behavior
 when the initial condition is in-phase though the oscillation behavior
 is a little complex in the case of the out-of-phase initial condition.
In the fermion transverse breathing oscillations, 
we found that the fermion-gas frequencies vary in time 
and converges into the same frequencies with the boson oscillations 
and the relative phases of the boson and fermion oscillations become in-phase. 
This phenomenon are shown to be explained 
by dampings of the intrinsic fermion modes when the boson motion is frozen.

In the case of the strongly-attractive boson-fermion interaction, 
$h_{BF}/g_{BB} \lesssim 2$,
the expansion of  fermi gas occurs in the transverse direction.
When the BF coupling is attractively large, most of fermions populate
inside the boson region of the ground state.
The boson gas oscillation heats the fermi gas and run the fermion out
from the boson occupied region, so that the fermi gas is expanded.
Such a behavior is
beyond that in the linear response theory and also beyond our approach.

These results show that
the longitudinal breathing oscillations in the largely deformed trap 
are similar with the quadrupole ones in the spherical trap  \cite{QPBF}.  
In contrast, 
the transverse breathing oscillations 
correspond to the monopole ones in the two dimensional system;
it is the similar property of the monopole oscillation 
in the three dimensional system. 

In this paper we discuss  the collective oscillations 
of the BF mixtures of Yb isotopes at $T=0$. 
The actual experiments have been done at finite temperatures 
\cite{Fukuhara3,Fukuhara-M}.
In addition we need to examine the heating of the bose gas as mentioned above. 
Thus thermal bosons should give some contributions;
the introduction of such effects into the present approach 
through two-body collision terms \cite{JackZar,BUU1} 
should be done also in the future.

\bigskip
{\bf Acknowledgement}

This work is supported in part by the Japanese Grand-in-Aid for
Scientific Research Fund of the Ministry of Education, Science, Sports
and Culture (21540212 and 22540414).

\end{document}